%% file: paper.tex
\documentclass[11pt]{article}

\usepackage[a4paper,margin=2.5cm]{geometry}
\usepackage[auth-sc-lg,affil-sl]{authblk}
\usepackage{natbib}
\usepackage[algoruled,vlined]{algorithm2e}
\usepackage{amsmath}
\usepackage{xcolor}
\usepackage{graphicx}
\usepackage{nomencl}

\include{format}

\newcommand{\save}[1]{}

\newcommand{\elew}{_{w,e}}

\title{Finite element model based on refined plate theories for laminated glass units}

\author[1]{Alena Zemanov\'{a}}
\author[1,2]{Jan Zeman}
\author[1]{Michal \v{S}ejnoha}
\affil[1]{Department of Mechanics, Faculty of Civil Engineering,
  Czech Technical University in Prague, Th\'{a}kurova 7, 166 29 Prague
  6, Czech Republic}
\affil[2]{Centre of Excellence IT4Innovations, V\v{S}B-TU Ostrava, 
17.~listopadu 15/2172 708 33 Ostrava-Poruba, Czech Republic}

\date{}

\begin{document}
\baselineskip=16pt
\maketitle

\begin{abstract} 
\baselineskip=16pt
Laminated glass units exhibit complex response as a result of different mechanical behavior and properties of glass and polymer foil. We aim to develop a finite element model for elastic laminated glass plates based on the refined plate theory by Mau. For a geometrically nonlinear description of the behavior of units, each layer behaves according to the Reissner-Mindlin kinematics, complemented with membrane effects and the von K\'{a}rm\'{a}n assumptions. Nodal Lagrange multipliers enforce the compatibility of independent layers in this approach. We have derived the discretized model by the energy-minimization arguments, assuming that the unknown fields are approximated by bi-linear functions at the element level, and solved the resulting system by the Newton method with consistent linearization. We have demonstrated through verification and validation examples that the proposed formulation is reliable and accurately reproduces the behavior of laminated glass units. This study represents a first step to the development of a comprehensive, mechanics-based model for laminated glass systems that is suitable for implementation in common engineering finite element solvers.
\end{abstract}


\section{Introduction}\label{sec:introduction}
%%%%%%%%%%%%%%%%%%%%%%%%%%%%%%%%%%%%%%%%%%%%%%%%%%%%%%%%%%%%%%%%%%%%%%%%%%%%%%%%%%%%%%%%%%%%%%%%%%%%%%%%%%%%%%%%%%%%%%%%%%%%%%%%%%

% Seznameni s pojmem laminovane sklo
%Glass is a popular and attractive material for transparent structures. The use of glass in buildings started with traditional window panes and has expanded to large facades, roof and floor systems, or staircases. Such architectural trend requires large-area glass panes and this is facilitated by the development of new material systems. One of such improved products is laminated glass formed by glass panes and polymer interlayers made of, for example, Polyvinylbutyral (PVB). Combination of brittle glass plate and ductile polymer foil improves the post-breakage behavior, which is a reason for using laminated glass panes for load-bearing glass structures or in situations when the glass could fall if shattered. 

Laminated glass has been developed to increase safety and security of fragile thin glass sheets. 
% Prehled literatury a odkaz na nosniky
When the load is transferred in both in-plane directions, the unit cannot be considered as a beam and a more accurate description is needed. Thus, the goal of this paper is an extension of our earlier finite element model of laminated glass beams~\citep{Zemanova:2014:NMFS} towards laminated glass plates. For this reason, the overview of available literature below will focus on laminated glass plates, whereas the discussion related to beams can be found in~\cite{Zemanova:2014:NMFS}. 
% Shrnuti bodu dulezitych pro modelovani
Let us summarize that the laminated glass units behave in an unusual manner due to several reasons. First, the polymer foil has the elastic modulus which is by orders of magnitude smaller than the elastic modulus of the glass. Such \textit{heterogeneity} in elastic properties of individual layers complicate the modeling of laminated glass units. Second, the polymer foil is highly \textit{viscoelastic with a great temperature dependency}. Third, \textit{geometric nonlinearity} should be included in the laminated glass behavior, which is caused by the small thickness of plates compared to the other dimensions.

% Vyber limitniho stavu
A number of experimental measurements, e.g.~\citep{Vallabhan:1992:PPVB,Behr:1993:SBA,Bennison:1999:FLB}, studied influence of temperature and load duration on a structural behavior of a laminated glass unit since the polymer foil is a time/temperature-dependent material, whose performance determines the coupling along glass layers.
% Dhaliwal:2002:CPVBTA,
The basic method to approximate the mechanical response of laminated glass units involves the utilization of one of the limiting cases: layered case as an assembly of independent glass layers and monolithic case with thickness equal to the combined thickness of glass layers and interlayers\footnote{This formulation corresponds to the true monolithic model, whereas the equivalent monolithic model has the thickness equal only to the total thickness of glass layers. For the upper bound it is more correct to assume the true monolithic model.
} determined under the assumption of geometric linearity.\footnote{Note that the interval between the monolithic and layered limit can be reduced by including the effects of geometric nonlinearity, see~\cite{Vallabhan:1987:SLG}.} In~\cite{Behr:1985:LGU}, the recommendation for the choice of the limiting case can be found, which takes into account only the temperature dependence, whereas the extension of the previous experimental stress analysis by load duration influence is summarized in~\cite{Behr:1993:SBA}.
Several approaches to the more precise analytical or numerical analysis of laminated glass structures have been proposed in order to ensure a reliable and efficient design.

% Efektivni tloustka
Single-layer approaches, useful especially in the preliminary phases of the design procedure, approximate the behavior of unit by an equivalent homogeneous plate and can be found in a draft version of European standard for glass~\cite{prEN16612}. \cite{Benninson:2008:HPLG} extended the concept of effective thickness to the case of laminated glass beams and determined the thickness of equivalent monolithic glass unit as a function of load duration and temperature history. 
\cite{Galuppi:2012:ETLGP} proposed new relations for the deflection- and stress-effective thicknesses of geometrically linear laminated glass plates using variational principles and approximation of shape function for deformation of laminated glass plate. Practical expressions for the design of laminated glass was summarized in~\cite{Galuppi:2012:PEFD}. The effects of geometric nonlinearity are not considered in this approach although they can be significant for laminated glass units with a small value of the shear modulus of polymer foil. 

% Analyticke ci semianalyticke
Analytical methods provide closed-form solutions only for laminated beams with very specific boundary conditions and when excluding the effects of geometric nonlinearity; for plates only series-based solutions are available.
\cite{Foraboschi:2012:AMLGP} described a mechanical behavior of laminated glass plates by a system of differential equations derived under the Kirchhoff assumptions for two thin glass plates with the same thickness. He considered only shear stresses and strains in the polymer interlayer and solve the resulting system by the Fourier series expansion for the load and displacements. His approach is valid for a simply supported rectangular laminated glass plate subjected to lateral uniformly distributed static loading. An extension of this model to other sandwich plates with stiff interlayer can be found in~\cite{Foraboschi:2013:TLPES}.
% Kdy nelze pouzit
These analytical approaches are rather difficult to be generalized to units with multiple layers and are mainly useful for the verification of numerical models. 

% Numericke - metoda siti
We find the numerical analysis of the discretized formulation of the problem as the most advantageous for the modeling of nonlinear behavior. The most common tool for structural analysis is the finite element method, nevertheless a few approaches based on finite difference method can be found in literature, e.g. \cite{Vallabhan:1993:ALG} and \cite{Asik:2003:LGP}.
% Numericke konecneprvkove
%
The first possibility for the simulation of behavior of laminated glass units utilizes three-dimensional solid elements for individual layers and their interaction, cf.~\citep{Duser:1999:AGBL}. This approach requires a large number of solid finite elements and therefore leads to expensive computations, since the thickness of the laminates, especially the polymer foil, is small compared to the other dimensions. 
To avoid fully resolved three-dimensional simulations, we used a refined plate theory in our approach and considered the layer-wise linear variation of in-plane displacements in the thickness direction. Most of the commercial finite element codes do not have such a feature, but rely on classical laminate formulations. In this framework, the laminated glass is modeled as a homogeneous unit with effective stiffness determined from its layout and standard plate (shell) elements are employed in the analysis. Because of a great mismatch of the material properties of PVB and glass, this approach is not appropriate as it is too coarse to correctly represent the inter-layer interactions. 

% Cleneni clanku
%In this paper, a derivation of the numerical model follows this path: starting with the mechanical model, Section~\ref{sec:formulation}, that is discretized with finite elements in order to arrive at the final system of linear equations in Section~\ref{sec:implementation}. A detailed derivation of stiffness matrices needed for implementation is presented in~\cite{Zemanova:2014:NMLG}. % odkaz na tezi  Examples for verification and validation of the finite element model and the discussion of results can be found in Section~\ref{sec:examples}.  Conclusions are summarized in Section~\ref{sec:conclusions}.

\section{Model assumptions}\label{sec:assumptions}
%%%%%%%%%%%%%%%%%%%%%%%%%%%%%%%%%%%%%%%%%%%%%%%%%%%%%%%%%%%%%%%%%%%%%%%%%%%%%%%%%%%%%%%%%%%%%%%%%%%%%%%%%%%%%%%%%%%%%%%%%%%%%%%%%%

% Navrzeny model
% 1. Jak se vyporada s heterogenitou vrstev / zlomem
The proposed numerical model for laminated glass plates is inspired by a specific class of refined plate theories~\citep{Mau:1973:RLP,Sejnoha:1996:MMU,Kruis:2002:SLP}. In this framework, each layer is treated as a shear deformable plate with independent kinematics. Interaction between individual layers is captured by the Lagrange multipliers (with a physical meaning of nodal forces), which result from the fact that no slippage is supposed to occur between glass plate and polymer foil. % Zminit, ze je to ale mozne

% 2. Jak se vyporada s viscoelasticitou a vyhled do budoucna
As for constitutive modeling, glass typically responds in a linear elastic manner, whereas the polymer foil behaves in a viscoelastic manner.  However, in many cases even the foil is modeled as being linear elastic by means of an effective shear modulus related to temperature and duration of loading. The hypothesis that the viscoelastic material is approximated by the linear elastic one, assuming a known temperature and duration of loading, may be constraining. On the other hand, a viscoelastic model of interlayer is more difficult to manage, and will be discussed independently and in more details in forthcoming publications.

% Jak zohlednujeme geometrickou nelinearitu
As for the effects of geometric nonlinearity, we have reported that they should not be neglected when modeling laminated glass structures, and also confirmed this statement by numerical examples in~\cite{Zemanova:2014:NMFS}, based on the fully nonlinear Reissner \emph{beam} formulation. Before extending the model to the laminated plates, we compared first the response of the fully nonlinear formulation with a simpler model based on the von K\'{a}rm\'{a}n assumptions, in which only deflections are large whereas the in-plane displacements and rotations remain small. Since the response of these two models was found to be very similar, see~\cite{Zemanova:2014:NMLG}[Section 7.1], our strategy will be to accommodate the nonlinear effects due to vertical deflections only in membrane strains. 
% Zminit, ze je vhodny i na vicevrstva laminovana skla a odkaz na hybridni / [Overend, 2013]

% Znaceni
The following nomenclature is used in the text. Scalar quantities are denoted by lightface letters, e.g. $a$, and the bold letters are reserved for matrices, e.g. $\M{a}$ or $\M{A}$. $\M{A}\trn$ standardly stands for the matrix transpose and $\M{A}^{-1}$ for the matrix inverse. The subscript in parentheses, e.g. $a\lay{i}$, is used to emphasize that the variable $a$ is associated with the $i$-th layer.

\section{Model formulation}\label{sec:formulation}
%%%%%%%%%%%%%%%%%%%%%%%%%%%%%%%%%%%%%%%%%%%%%%%%%%%%%%%%%%%%%%%%%%%%%%%%%%%%%%%%%%%%%%%%%%%%%%%%%%%%%%%%%%%%%%%%%%%%%%%%%%%%%%%%%%

\subsection{Kinematics}\label{sec:kinematics}
%%%%%%%%%%%%%%%%%%%%%%%%%%%%%%%%%%%%%%%%%%%%%%%%%%%%%%%%%%%%%%%%%%%%%%%%%%%%%%%%%%%%%%%%%%%%%%%%%%%%%%%%%%%%%%%%%%%%%%%%%%%%%%%%%%
Being inspired by~\cite{Pica:1980:FEA}, each layer is modeled as the Reissner-Mindlin plate with nonlinear membrane effects, by adopting
the following kinematic assumptions:
\begin{itemize}
\item a straight line segment, normal to the underformed mid-surface, remains
straight but not necessarily normal to the deformed mid-surface,
\item vertical displacements do not vary along the thickness of the plate.
\end{itemize}
\begin{figure}[ht]
\centerline{%
 \includegraphics[height=70mm]{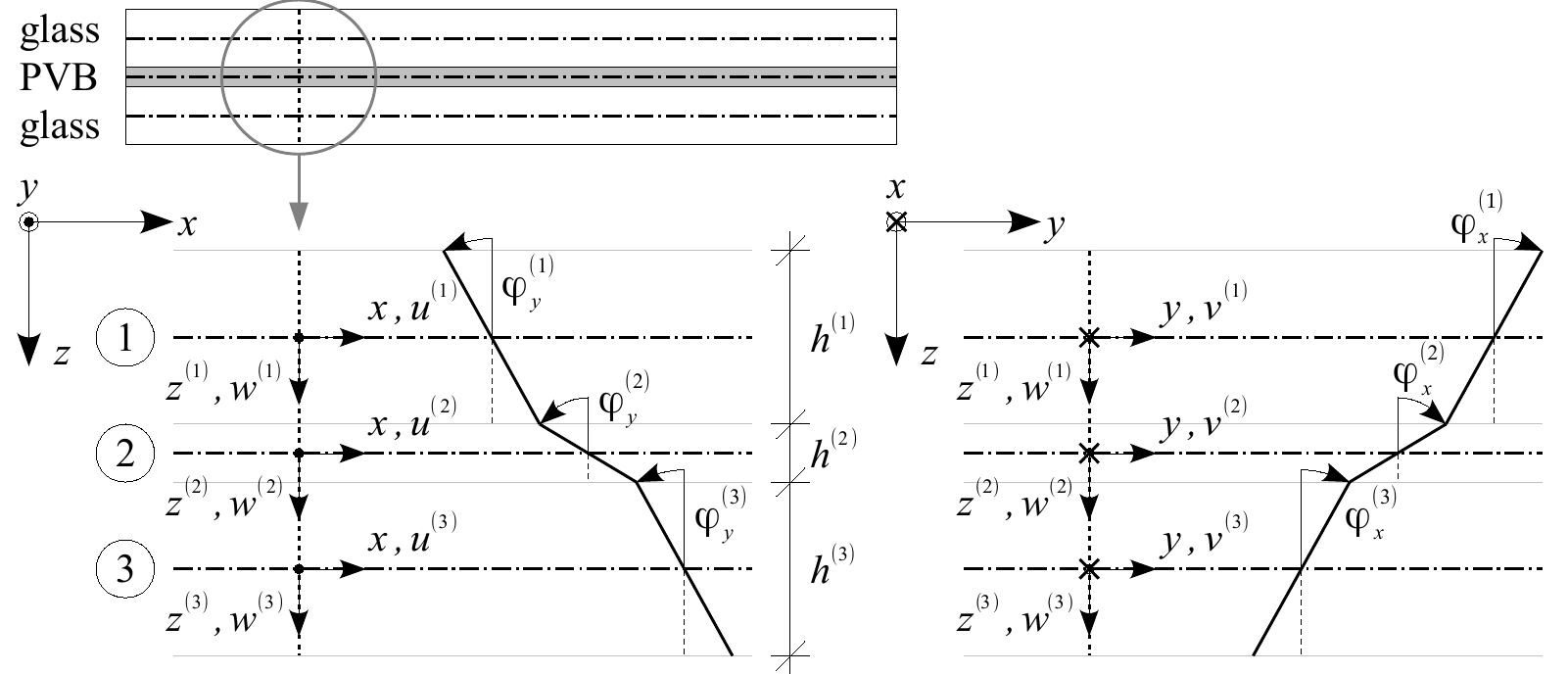} 
}%
\caption{Kinematics of laminated plate assuming the layer-wise linear variation of in-plane displacements in the thickness direction}
\label{fig:lam_plate}
\end{figure}
Thus, the nonzero displacement components in each layer $\dispx\lay{i}$, $\dispy\lay{i}$, and $\dispz\lay{i}$
are parametrized as, see Figure~\ref{fig:lam_plate},
\begin{subequations}
\label{eq4:displac}
\begin{eqnarray}
u\layer{i}(x,y,z\layer{i}) &=& u_0\layer{i}(x,y) + \roty\layer{i}(x,y) z\layer{i}, \\
v\layer{i}(x,y,z\layer{i}) &=& v_0\layer{i}(x,y) - \rotx\layer{i}(x,y) z\layer{i}, \\
w\layer{i}(x,y,z\layer{i}) &=& w_0\layer{i}(x,y),
\end{eqnarray}
\end{subequations}
where $i=1,2,3$ refers to individual layers, $u_0$, $v_0$, and $w_0$ denote the
displacements of the mid-surface, $\rotx$ and $\roty$ represent the straight line segment
rotations about the corresponding axes, and $z\layer{i}$ is measured in
the local coordinate system relatively to the mid-plane of the $i$-th layer.
For the layer surfaces to be connected, the corresponding displacements must satisfy the
compatibility conditions
\begin{subequations}\label{eq4:continuityLEP}
\begin{eqnarray}
u\layer{i}(x,y,\frac{h\layer{i}}{2}) - u\layer{i+1}(x,y,-\frac{h\layer{i+1}}{2}) &=& 0,\\
v\layer{i}(x,y,\frac{h\layer{i}}{2}) - v\layer{i+1}(x,y,-\frac{h\layer{i+1}}{2}) &=& 0,\\
w\layer{i}(x,y,\frac{h\layer{i}}{2}) - w\layer{i+1}(x,y,-\frac{h\layer{i+1}}{2}) &=& 0,
\end{eqnarray}
\end{subequations}
with $i=1,2$.

The nonzero components of the Green-Lagrange strain tensor, e.g.~\cite[Section~24.1]{Jirasek:2002:IAS}, at the $i$-th layer can be divided into the membrane effects 
\begin{equation}
\mtrx{\strain}\m\layer{i}(x,y,z\layer{i}) = \left[ \strain\layer{i}_{x} (x,y,z\layer{i}), \strain\layer{i}_{y} (x,y,z\layer{i}), \sstrain\layer{i}_{xy} (x,y,z\layer{i}) \right]\trn
\end{equation}
and the transverse shear strains 
\begin{equation}
\mtrx{\sstrain}\layer{i}(x,y) = \left[ \sstrain\layer{i}_{xz}(x,y), \sstrain\layer{i}_{yz}(x,y) \right]\trn.
\end{equation}
Under the von K\'{a}rm\'{a}n assumptions~\citep{Timoshenko:1959:TPS}, the derivatives of $u\layer{i}$ and $v\layer{i}$ with respect to all coordinates are small and  
these parts of strain tensor become
\begin{subequations}
\begin{align}
\mtrx{\strain}\m\layer{i}(x,y,z\layer{i}) 
& =  
\pard\mtrx{u}\layer{i}_0(x,y) + \pard\mtrx{S}\mtrx{\rot}\layer{i}(x,y) z\layer{i} + 
\mtrx{\strain}\mONL\layer{i}(x,y)\nonumber\\
& =
\mtrx{\strain}\mO\layer{i}(x,y) 
+
\mtrx{\curv}\layer{i}(x,y) z\layer{i}
+ 
\mtrx{\strain}\mONL\layer{i}(x,y),
\label{eq:generalized_strains_a}\\
\mtrx{\sstrain}\layer{i}(x,y) 
& = 
\mtrx{\nabla}w_0\layer{i}(x,y) 
+ 
\mtrx{S}
\mtrx{\rot}\layer{i}(x,y),
\label{eq:generalized_strains_b}
\end{align}
\label{eq:generalized_strains}
\end{subequations}
and are expressed in terms of the mid-plane displacements, rotations, and pseudo-curvatures as
\begin{subequations}
\begin{align}
\mtrx{u}\layer{i}_0(x,y) &= \left[{u}\layer{i}_0(x,y), {v}\layer{i}_0(x,y) \right]\trn,\\
\mtrx{\rot}\layer{i}(x,y) &= \left[\rot\layer{i}_x(x,y), \rot\layer{i}_y(x,y)\right]\trn,\\
\mtrx{\curv}\layer{i}(x,y) &= \left[ \curvx\layer{i}(x,y), \curvy\layer{i}(x,y), \curvxy\layer{i}(x,y) \right]\trn.
\end{align}
\end{subequations}
Observe that the shear strain components and pseudo-curvatures are the same as in the geometrically linear case, but the membrane strain components include nonlinear
contributions depending on the deflections of the layer mid-surface
\begin{align}
\mtrx{\strain}\mONL\layer{i}(x,y) =\left[\frac{1}{2} \left( \ppd{w_0\layer{i}}{x}(x,y)\right)^2, \frac{1}{2} \left( \ppd{w_0\layer{i}}{y}(x,y)\right)^2, \ppd{w_0\layer{i}}{x}(x,y) \ppd{w_0\layer{i}}{y}(x,y)\right]\trn.
\label{eq:strain_NL}
\end{align}
The matrix $\mtrx{S}$ and the differential operators $\pard$ and
$\mtrx{\nabla}$ are provided by
\begin{align}
\pard = 
	\left[
	\begin{array}{cc}
	\ppd{}{x} & 0 \\ 0 & \ppd{}{y} \\ \ppd{}{y} & \ppd{}{x}
	\end{array}
	\right],
&&
\mtrx{\nabla} =
	\left[
	\begin{array}{c}
\ppd{}{x} \\
\ppd{}{y}
	\end{array}
   \right],
&&
\mtrx{S} =
	\left[
	\begin{array}{cc}
	0 & 1 \\ -1 & 0
	\end{array}
	\right].
\end{align}

\subsection{Constitutive relations and specific internal forces}\label{sec:constitutive}
%%%%%%%%%%%%%%%%%%%%%%%%%%%%%%%%%%%%%%%%%%%%%%%%%%%%%%%%%%%%%%%%%%%%%%%%%%%%%%%%%%%%%%%%%%%%%%%%%%%%%%%%%%%%%%%%%%%%%%%%%%%%%%%%%%
Having parametrized the strains with quantities defined on the mid-surface, we
proceed to stresses. We assume that the normal stress in the $z$ direction is negligible compared to the other stresses. It follows from the generalized Hooke law for isotropic
bodies that the membrane stress components and the shear stress components are given by, cf.~\citep{Pica:1980:FEA},
\begin{align}\label{eq5:stress_strain_membrane}
\mtrx{\stress}\layer{i}_{m} (x,y,z\layer{i}) = \Dm\layer{i} \mtrx{\strain}\m\layer{i}(x,y,z\layer{i}),
&&
\mtrx{\sstress}\layer{i} (x,y) = G\layer{i} \I \mtrx{\sstrain}\layer{i}(x,y),
\end{align}
where
\begin{align}
\Dm\layer{i} = \frac{E\layer{i}}{1-(\nu\layer{i})^2}
\left[
	\begin{array}{ccc}
	1 & \nu\layer{i} & 0\\
	\nu\layer{i} & 1 & 0\\
	0 & 0 & \displaystyle{\frac{1-\nu\layer{i}}{2}}
	\end{array}
\right],
\end{align}
is the membrane stiffness matrix, $E\layer{i}$ and $G\layer{i}$ are Young's modulus of elasticity and shear modulus of the $i$-th layer, and $\I$ is the identity matrix. Notice that, since the Reissner-Mindlin kinematics were utilized, the transverse shear stresses are approximated as constant in the $z\lay{i}$ direction.%, in comparison to the equilibrium quadratic profile.

The specific internal forces energetically conjugate to the generalized strains~\eqref{eq:generalized_strains}, are obtained by suitable weighted stress averages in the thickness direction. As for the membrane strains $\mtrx{\strain}\m\layer{i}$, these are associated with the specific normal forces $\mtrx{\normal}\lay{i}$ in the form
\begin{align}
\mtrx{\normal}\layer{i}(x,y)
&
=
[\normal_x\layer{i}(x,y), \normal_y\layer{i}(x,y), \normal_{xy}\layer{i}(x,y)]\trn
=  
\int_{-\frac{\h\layer{i}}{2}}^{\frac{\h\layer{i}}{2}}
{\mtrx{\stress}\m\layer{i}(x,y,z\layer{i})}\de z\layer{i} 
\nonumber
\\
& = 
\h\layer{i} \Dm\layer{i} 
\left( 
  \mtrx{\strain}\mO\layer{i}(x,y) 
  +
  \mtrx{\strain}\mONL\layer{i}(x,y) 
\right) 
= 
\Dn\layer{i}
\left( 
  \pard\mtrx{u}\layer{i}_0(x,y) 
  + 
  \mtrx{\strain}\mONL\layer{i}(x,y) 
\right).
\label{eq4:specific_normal}
\end{align}
The quantity corresponding to the pseudo-curvature $\mtrx{\curv}\layer{i}$ is
the matrix of specific bending moments $\mtrx{\bendy}\layer{i}$ provided by
\begin{eqnarray}
\mtrx{\bendy}\layer{i}(x,y)
	&=& [\bendy_x\layer{i}(x,y), \bendy_y\layer{i}(x,y), \bendy_{xy}\layer{i}(x,y)]\trn
	 =  \int_{-\frac{\h\layer{i}}{2}}^{\frac{\h\layer{i}}{2}}{\mtrx{\stress}\m\layer{i}(x,y,z\layer{i})z\layer{i}}\de z\layer{i}\nonumber \\
	&=& \frac{(\h\layer{i})^3}{12} \Dm\layer{i} \mtrx{\curv}\layer{i}(x,y) =
	\Db\layer{i} \pard\mtrx{S}\mtrx{\rot}\layer{i}(x,y).
	\label{eq4:specific_moment}
\end{eqnarray}
The specific shear forces $\mtrx{\shear}\lay{i}$, conjugate to the shear strains
$\mtrx{\sstrain}\lay{i}$, are obtained in an analogous way
\begin{eqnarray}
\mtrx{\shear}\layer{i}(x,y)
	&=& [\shear_x\layer{i}(x,y), \shear_y\layer{i}(x,y)]\trn
	 =  M\correct\layer{i} \int_{-\frac{\h\layer{i}}{2}}^{\frac{\h\layer{i}}{2}}{\mtrx{\sstress}\layer{i}(x,y)}\de z\layer{i} \nonumber\\
	&=&  \correct\layer{i} \G\layer{i} \I \h\layer{i} \mtrx{\sstrain}\layer{i}(x,y) 
	 = \Ds\layer{i} \left(\mtrx{\nabla}w_0\layer{i}(x,y) + \mtrx{S}
	 \mtrx{\rot}\layer{i}(x,y)\right),
	 \label{eq4:specific_shear}
\end{eqnarray}
but need to be corrected due to the assumed constant shear stresses by a constant $\correct\layer{1} = \correct\layer{3} =5/6$ for top and bottom layers and $\correct\layer{2} = 1$ for interlayer.\footnote{We assumed the constant value of 5/6 for glass layers, which corresponds to a parabolic distribution in the thickness direction with almost-zero values on top and bottom surfaces. However interlayer is supposed to exhibit a state of pure shear with constant distribution and therefore we set the value to 1.} In Eqs.~\eqref{eq4:specific_normal}--\eqref{eq4:specific_shear}, $\Dn\lay{i}$, $\Db\lay{i}$, and $\Ds\lay{i}$ denote the matrices of normal, bending, and shear stiffness of the $i$-th layer, respectively.

Thus in the expressions for the specific internal forces, the geometric nonlinearity due to von K\'{a}rm\'{a}n assumptions appears in the normal
forces only, and the specific moments and shear forces remain in the same form as in the geometrically linear case. %The same holds for the internal energy of the $i$-th layer. 

\subsection{Energy functional}\label{sec:energy}
%%%%%%%%%%%%%%%%%%%%%%%%%%%%%%%%%%%%%%%%%%%%%%%%%%%%%%%%%%%%%%%%%%%%%%%%%%%%%%%%%%%%%%%%%%%%%%%%%%%%%%%%%%%%%%%%%%%%%%%%%%%%%%%%%%
The governing equations of the discretized model are derived by the energy-minimization arguments. To this purpose, we express the internal energy of the $i$-th layer in the form 
\begin{eqnarray}\label{eq5:EintLEP}
\Eint\layer{i}
	\left( 
		\mtrx{\strain}\layer{i}\mO, 
		\mtrx{\strain}\layer{i}\mONL, 
		\mtrx{\sstrain}\layer{i}, 
		\mtrx{\curv}\layer{i} 
	\right)
=&&
\frac{1}{2}
\int_{\dmn}
(\mtrx{\strain}\layer{i}\mO+\mtrx{\strain}\layer{i}\mONL)\trn 
\Dn\layer{i} 
(\mtrx{\strain}\layer{i}\mO+\mtrx{\strain}\layer{i}\mONL)\\ \nonumber
&&+
(\mtrx{\sstrain}\layer{i})\trn \Ds\layer{i} \mtrx{\sstrain}\layer{i} 
+
(\mtrx{\curv}\layer{i})\trn \Db\layer{i} \mtrx{\curv}\layer{i}
\de \dmn,
\end{eqnarray}
where $\Omega\lay{i}$ denotes the mid-surface of the $i$-th layer. The potential energy of external loading acting at the $i$-th layer reads
\begin{eqnarray}
\Eext\layer{i}
\left(
	\mtrx{u}_0\layer{i},
	w_0\layer{i},
	\mtrx{\rot}\layer{i} 
\right)
= 
& - &
\int_{\dmn}
\left(
	(\mtrx{u}_0\layer{i})\trn \overline{\mtrx{f}}\m\layer{i} 
	+
	w_0\layer{i} \overline{f}_z\layer{i} 
	+
	(\mtrx{\rot}\layer{i})\trn \overline{\mtrx{m}}\layer{i} 
\right)
\de \dmn,
\label{eq4:EextLEP}
\end{eqnarray}
with $\overline{\mtrx{f}}\m\layer{i}$, $\overline{f}_z\layer{i}$, and $\overline{\mtrx{m}}\layer{i}$ referring to the intensity of the distributed
in-plane, out-of-plane and moment loads. The total energy of the $i$-th layer is expressed as the sum of internal and external components
\begin{align}
\Etot\layer{i} 
\left(
	\mtrx{u}_0\layer{i},
	w_0\layer{i},
	\mtrx{\rot}\layer{i}
\right)
& =
\Eint\layer{i}
\left(
\pard\mtrx{u}\layer{i}_0 + \mtrx{\strain}\mONL\layer{i}(w_0\layer{i}),
\mtrx{\nabla}w_0\layer{i} + \mtrx{S} \mtrx{\rot}\layer{i},
\pard\mtrx{S}\mtrx{\rot}\layer{i}
\right)
\nonumber \\
& +
\Eext\layer{i} 
\left(
	\mtrx{u}_0\layer{i},
	w_0\layer{i},
	\mtrx{\rot}\layer{i}
\right),
\end{align}
in order to assemble the total energy of the system from the contributions of individual layers
\begin{equation}
\Etot
\left(
	\mtrx{u}_0\layer{1}, w_0\layer{1}, \mtrx{\rot}\layer{1},
	\mtrx{u}_0\layer{2}, w_0\layer{2}, \mtrx{\rot}\layer{2},
	\mtrx{u}_0\layer{3}, w_0\layer{3}, \mtrx{\rot}\layer{3}
\right)
=
\sum_{i=1}^3
\Etot\layer{i}
\left(
	\mtrx{u}_0\layer{i},
	w_0\layer{i},
	\mtrx{\rot}\layer{i}
\right).
\end{equation}
The true fields $\left( \mtrx{u}_0\layer{i}, w_0\layer{i}, \mtrx{\rot}\layer{i} \right)_{i=1}^{3}$ then follow from the minimization of functional
$\Etot$, subject to the prescribed boundary conditions and the compatibility conditions~\eqref{eq4:continuityLEP}.

\section{Numerical implementation}\label{sec:implementation} 
%%%%%%%%%%%%%%%%%%%%%%%%%%%%%%%%%%%%%%%%%%%%%%%%%%%%%%%%%%%%%%%%%%%%%%%%%%%%%%%%%%%%%%%%%%%%%%%%%%%%%%%%%%%%%%%%%%%%%%%%%%%%%%%%%%
% Finite element formulation \\ implementation % Numerical treatment

\subsection{Discretization}\label{sec:discretization}
%%%%%%%%%%%%%%%%%%%%%%%%%%%%%%%%%%%%%%%%%%%%%%%%%%%%%%%%%%%%%%%%%%%%%%%%%%%%%%%%%%%%%%%%%%%%%%%%%%%%%%%%%%%%%%%%%%%%%%%%%%%%%%%%%%
A standard discretization is employed, in which we approximate the unknown fields by bi-linear functions at the element
level, see Appendix~\ref{app:sensitivity_analysis}  % ahead to Section~\ref{app:sensitivity_analysis} 
for explicit expressions. 
%
%\begin{figure}[ht]
%\centerline{%
% \includegraphics[height=95mm]{figures/fin_elem_plate} 
%}%
%\caption{Finite element discretization}
%\label{fig:fin_elem_plate}
%\end{figure}
%
As a result of this step, the internal and external energies are approximated as 
\begin{align}
\Eint\layer{i}
	\left( 
		\mtrx{\strain}\layer{i}\mO, 
		\mtrx{\strain}\layer{i}\mONL, 
		\mtrx{\sstrain}\layer{i}, 
		\mtrx{\curv}\layer{i} 
	\right)
& \approx 
\sum_{e=1}^\nel
\Pi_{\mathrm{int},e}\layer{i}
\left( 
 \mtrx{r}\layer{i}\el 
\right),
\\
\Eext\layer{i} 
\left(
	\mtrx{u}_0\layer{i},
	w_0\layer{i},
	\mtrx{\rot}\layer{i}
\right)
& \approx 
\sum_{e=1}^\nel
\Pi_{\mathrm{ext},e}\layer{i}
\left( 
 \mtrx{r}\layer{i}\el 
\right)
=
-
\sum_{e=1}^\nel
 \mtrx{r}\layer{i}\el
 \trn 
 \Mfexte{i}{e}.
\end{align}
where $\mtrx{r}\layer{i}\el$ is the vector of generalized nodal displacements of
the $e$-th element and the $i$-th layer%, cf. Figure~\ref{fig:fin_elem_plate}
\begin{align}\label{eq4:r_el_LEP}
\mtrx{r}\layer{i}\el = \left[ 
\mtrx{r}\layer{i}\eln{1}, 
\mtrx{r}\layer{i}\eln{2}, 
\mtrx{r}\layer{i}\eln{3}, 
\mtrx{r}\layer{i}\eln{4}
\right]\trn,
&&
\mtrx{r}\layer{i}\eln{n} = \left[ 
u\layer{i}\ele{n}, v\layer{i}\ele{n}, w\layer{i}\ele{n}, \rot\layer{i}\elx{n}, \rot\layer{i}\ely{n}
\right]\trn,
\end{align}
and $\Mfexte{i}{e}$ stores the generalized nodal forces resulting from the distributed loading in~\eqref{eq4:EextLEP}, see again Appendix~\ref{app:sensitivity_analysis} for more details.

\subsection{Resulting system of governing equations}\label{sec:governing}
%%%%%%%%%%%%%%%%%%%%%%%%%%%%%%%%%%%%%%%%%%%%%%%%%%%%%%%%%%%%%%%%%%%%%%%%%%%%%%%%%%%%%%%%%%%%%%%%%%%%%%%%%%%%%%%%%%%%%%%%%%%%%%%%%%
In order to obtain the nodal degrees of freedom corresponding to external loading, we rely again on
variational arguments, and start with the discretized energy functional
\begin{equation}\label{eq:EintVK}
\Etot( \Md )
=
\sum_{i=1}^{3}
\sum_{e=1}^{\nel}
\Einte{i}{e}( \Mde{i}{e} )
+
\Eexte{i}{e}( \Mde{i}{e} )
\end{equation}
to be minimized under the kinematic constraints and compatibility
conditions. This produces the Lagrangian function in the form
\begin{equation}\label{eq:nonl_lagrangianVK}
\mathcal{L}( \Md, \M{\lambda} )
=
\Etot( \Md )
+
\M{\lambda}\trn\M{C}\Md,
\end{equation}
\nomenclature{$m$}{Index for Lagrange multipliers}%
where $\mtrx{r}$ stores arbitrary kinematically admissible generalized nodal displacements, and $\mtrx{\lambda}$ is the vector of admissible Lagrange
multipliers (with physical meaning of forces between layers due to compatibility).
The block of matrix $\mtrx{C}$, implementing the tying conditions between two adjacent layers at the $j$-th node, now assumes the form~($i=1,2$)
\begin{equation}
\begin{array}{c}
\mtrx{ C }\layer{i,i+1}_{j} =
\end{array}
\left[
 \begin{array}{ccccccccccc}
  1 & 0 & 0 &                     0 & \frac{h\layer{i}}{2} & ... & -1 &  0 &                       0 & 0 & \frac{h\layer{i+1}}{2} \\
  0 & 1 & 0 & -\frac{h\layer{i}}{2} &                    0 & ... &  0 & -1 &  0 & -\frac{h\layer{i+1}}{2}&                      0 \\
  0 & 0 & 1 &                     0 &                    0 & ... &  0 &  0 &                      -1 & 0 &                      0 \\
 \end{array}
\right],
\label{eq4:C1_LEP}
\end{equation}
so that the condition $\mtrx{C}\mtrx{r}=\mtrx{0}$
enforces the compatibility equations~\eqref{eq4:continuityLEP} discretely at all interfacial nodes. 
Notice that, due to the assumption that the rotations $\mtrx{\rot}$ remain small, the second term in~\eqref{eq:nonl_lagrangianVK} is linear in $\Md$, which significantly simplifies the derivation as well as implementation of the numerical model, when compared to the formulation presented in~\cite{Zemanova:2014:NMFS}. 

The Karush-Kuhn-Tucker optimality conditions, e.g.~\cite[Chapter~14]{Bonnans:2003:NOTPA}, associated with~\eqref{eq:nonl_lagrangianVK} read
\begin{subequations}\label{eq:opt_conditionsVK}
\begin{align}
\nabla_{\Md} 
\mathcal{L}\left( 
 \Md^*, \M{\lambda}^* 
\right)
=  
\nabla \Etot( \Md^* )
+ 
\M{C}\trn 	
\M{\lambda}^*
& =  
\M{0},
\label{eq:opt_conditions_1VK}
\\
\nabla_{\M{\lambda}} 
\mathcal{L}\left( 
 \Md^*, \M{\lambda}^* 
\right)
=
\M{C}\Md^*
& =  
\M{0},
\label{eq:opt_conditions_2VK}
\end{align} 
\end{subequations}
where $\Md^*$ and $\M{\lambda}^* $ are vectors of true nodal degrees of freedom and Lagrange multipliers. 
Due to the presence of quadratic terms in the normal strain~\eqref{eq:generalized_strains_a}, the
resulting system of nonlinear equations must be resolved by the Newton scheme, by
iteratively searching for the correction $\ite{\step+1}\delta\Md$ to the known
nodal displacements $\ite{\step}\Md$ in the form
\begin{equation}\label{eq:r_deltarVK}
\ite{\step+1}\Md
=
\ite{\step}\Md
+
\ite{\step+1}\delta\Md.
\end{equation}
To this purpose, we linearize the term
\begin{align}
\nabla\Etot( \ite{\step+1}\Md ) 
& \approx 
\nabla\Etot( \ite{\step}\Md )
+
\nabla^2 \Etot( \ite{\step}\Md)
\ite{\step+1}\delta\Md,
\end{align}
and plug it into the optimality condition~\eqref{eq:opt_conditions_1VK}. As a
result, we receive a linear system in the form
\begin{equation}\label{eq:systemVK}
\begin{bmatrix}
 \ite{\step}\M{K} & \M{C}\trn \\
 \M{C} & \M{0}
\end{bmatrix}
\begin{bmatrix}
 \ite{\step+1} \delta \Md \\
 \ite{\step+1} \M{\lambda}
\end{bmatrix}
=
-
\begin{bmatrix}
 \ite{\step}\Mfint - \Mfext \\
 \M{0}
\end{bmatrix}
,
\end{equation}
in which 
\begin{subequations}
\begin{align}
\ite{\step}\M{K} 
& =  
\nabla^2 \Etot( \ite{\step}\Md )
=
\MKt( \ite{\step}\Md ),
\\
\ite{\step}\Mfint - \Mfext
& =  
\nabla \Etot( \ite{\step}\Md ).
\end{align}
\end{subequations}
Due to the fact that each layer is treated separately, the matrices appearing in
Eq.~\eqref{eq:systemVK} possess the block structure
\begin{equation}\label{eq:matrices_block_diagonalVK}
\MKt( \Md )
=
\begin{bmatrix}
\MKt\lay{1}( \Md\lay{1} ) \\ 
& 
\MKt\lay{2}( \Md\lay{2} ) \\
&&  
\MKt\lay{3}( \Md\lay{3} )
\end{bmatrix}
, \quad
\Mfint
=
\begin{bmatrix}
 \Mfint\lay{1} \\
 \Mfint\lay{2} \\
 \Mfint\lay{3}
\end{bmatrix}
, \quad
\Mfext
=
\begin{bmatrix}
 \Mfext\lay{1} \\
 \Mfext\lay{2} \\
 \Mfext\lay{3} 
\end{bmatrix},
\end{equation}
and are obtained by the assembly of the contributions of individual elements, cf.~\citep{Bittnar:1996:NMM}  
\begin{align}\label{eq:internal_forcesVK}
\M{f}_{\mathrm{int},e}\lay{i}
=
\ppd{\Etot_{\mathrm{int},e}\lay{i}}{\Mde{i}{\el}}, 
&&
\M{K}_{e}\lay{i} 
=
\ppd{^2\Etot_{\mathrm{int},e}\lay{i}}{\Mde{i}{\el}{}^2} 
=
\ppd{\Mfinte{\el}\lay{i}}{\Mde{i}{\el}}.
\end{align} 
The explicit expression for the internal forces and the tangent matrices are
elaborated in detail in Appendix~\ref{app:sensitivity_analysis}. 

\subsection{Implementation issues}\label{sec:algorithm}
%%%%%%%%%%%%%%%%%%%%%%%%%%%%%%%%%%%%%%%%%%%%%%%%%%%%%%%%%%%%%%%%%%%%%%%%%%%%%%%%%%%%%%%%%%%%%%%%%%%%%%%%%%%%%%%%%%%%%%%%%%%%%%%%%%
The resulting version of the algorithm is shown in Algorithm~\ref{alg:impl_NRVK}, where the convergence of the iteration is checked by the equilibrium residual.
\begin{align}\label{eq:residualsVK}
\ite{\step}\eta
= 
\frac{%
\| \ite{\step}\Mfint - \Mfext + \M{C}\trn
\ite{\step}\M{\lambda} \|_2
}{%
\max \left(\| \Mfext \|_2, 1 \right)
}.
\end{align}
\begin{algorithm}[ht]
 \KwData{initial displacement $\ite{0}\Md$, tolerance $\epsilon$}
 \KwResult{$\Md^*$, $\M{\lambda}^*$}
 $\step \leftarrow 0, \ite{0}\M{\lambda} \leftarrow \M{0}$, 
 assemble $\ite{\step}\Mfint$ and $\M{C}$\\ 
 \While{$(\ite{\step}\eta > \epsilon)$}{%
 assemble $\ite{\step}\M{K}$ \\
 solve for $(\ite{\step+1}\delta \Md, \ite{\step+1}\M{\lambda})$ from \Eref{eq:systemVK}\\ 
 $\ite{\step+1} \Md \leftarrow \ite{\step}\Md + \ite{\step+1}\delta \Md$\\
 assemble $\ite{\step+1}\Mfint$\\ 
 $\step\leftarrow \step+1$ \\
 } 
 $\Md^* \leftarrow \ite{\step}\Md$, $\M{\lambda}^* \leftarrow \ite{\step}
 \M{\lambda}$
 \caption{Conceptual implementation of the Newton method for von
 K\'{a}rm\'{a}n plate}
 \label{alg:impl_NRVK}
\end{algorithm}

Note that the finite element model of geometrically nonlinear laminated glass plate was implemented in MATLAB-based application LaPla. In the post-processing step, the stress components are evaluated from the discretized constitutive equations~\eqref{eq5:stress_strain_membrane} at the element level and extrapolated to nodes using standard smoothing procedures, e.g.~\citep{Hinton:1974:LGS}. An interested reader is referred to~\cite{Zemanova:2014:NMLG} for a brief description of MATLAB-based program LaPla.

\section{Examples and discussion}\label{sec:examples} 
%%%%%%%%%%%%%%%%%%%%%%%%%%%%%%%%%%%%%%%%%%%%%%%%%%%%%%%%%%%%%%%%%%%%%%%%%%%%%%%%%%%%%%%%%%%%%%%%%%%%%%%%%%%%%%%%%%%%%%%%%%%%%%%%%%

We validated and verified the developed finite element model, Section~\ref{sec:validation}, and compared it against the semi-analytical and effective thickness solutions, Sections~\ref{sec:verAM} and~\ref{sec:verET}. 

\subsection{Experimental validation and numerical verification}\label{sec:validation} 
%%%%%%%%%%%%%%%%%%%%%%%%%%%%%%%%%%%%%%%%%%%%%%%%%%%%%%%%%%%%%%%%%%%%%%%%%%%%%%%%%%%%%%%%%%%%%%%%%%%%%%%%%%%%%%%%%%%%%%%%%%%%%%%%%%

%The geometrically nonlinear model of elastic laminated glass plate is validated against the data by~\cite{Vallabhan:1993:ALG}, corresponding to the results of the experiments done in Glass Research and Testing Laboratory at Texas Tech University, Lubbock, Texas, USA. In addition, we also verified our results against the finite difference model of~\cite{Asik:2003:LGP}.

We considered a laminated glass unit simply supported along all four edges as an example problem for the deflection and stress analysis. \cite{Vallabhan:1993:ALG} performed a series of tests on such laminated units. In-plane dimensions of the unit were 1.5~m $\times$ 1.5~m and the unit was subjected to a transverse loading,  gradually increasing to 6.9~kPa in 60~s. The temperature during a series of tests varied between 21$\,^\circ$C and 27$\,^\circ$C. The maximum lateral displacements and strains at several locations at the top and bottom of the plate were measured and converted to the principal stresses.

\Tref{Tab:properties} summarizes the thicknesses and material properties of individual layers. The material parameters of glass were taken from~\cite{Vallabhan:1993:ALG}, because these values were assumed for the conversion of measured strains to stresses. The Poisson ratio for interlayer is according to draft European standard by~\cite{prEN16613}. The value of the shear modulus of PVB was found by matching the measured central deflection of laminated glass units with the values obtained by the numerical models. \cite{Vallabhan:1993:ALG} obtained an agreement between experiment and finite-difference solver using the value of 689.5~kPa in their study, whereas we matched experimental data with a smaller value of 400~kPa. Our finite element model is stiffer due to the explicit inclusion of interlayer thickness and this fact is in accordance with the results of the fully three-dimensional finite element analysis by~\cite{Duser:1999:AGBL}, and so is the decreased optimal value of the shear modulus.

\begin{table}[ht]
\caption{Properties of glass and PVB layers}
\centering{
\begin{tabular}{lcc}
\hline
\textbf{Layer/material} & \textbf{Glass} ($i=1,3$) & \textbf{PVB} ($i=2$)\\
\hline
Thickness $t\layer{i}$ [mm] & 4.76 & 1.52 \\
Young's modulus of elasticity $\E\layer{i}$ [MPa] & 68,900 & --\\
Shear modulus of elasticity $\G\layer{i}$ [MPa] & -- & 0.4 / 0.6895 \\
Poisson's ratio $\nu\layer{i}$ [-] & 0.22 & 0.49\\
\hline
\end{tabular}
}
\label{Tab:properties}
\end{table}

\begin{figure}[ht]
\centerline{%
 \includegraphics[width=100mm]{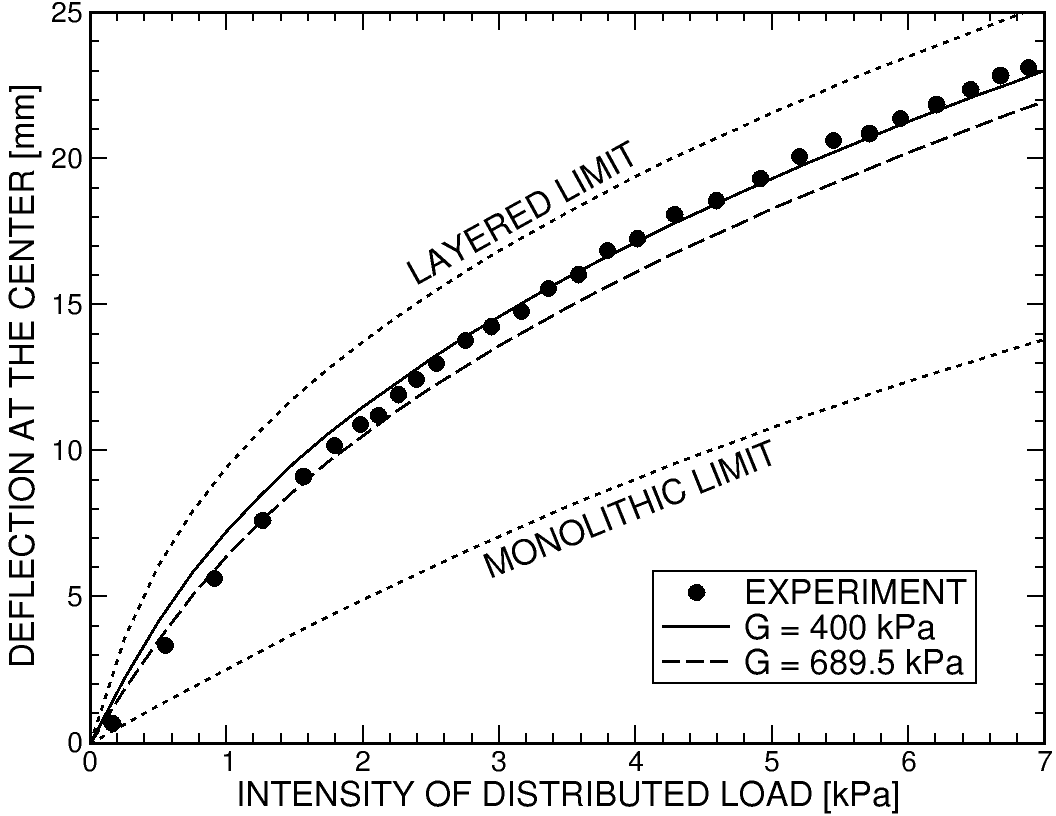} 
}%
\caption{Comparison of experimental data for central deflection with the response of nonlinear finite element model of laminated glass plate with the PVB shear modulus 400~kPa or 689.5~kPa}
\label{fig:Exp_w}
\end{figure}

\begin{figure}[p]
\centerline{%
\begin{tabular}{cc}
 \includegraphics[width=75mm]{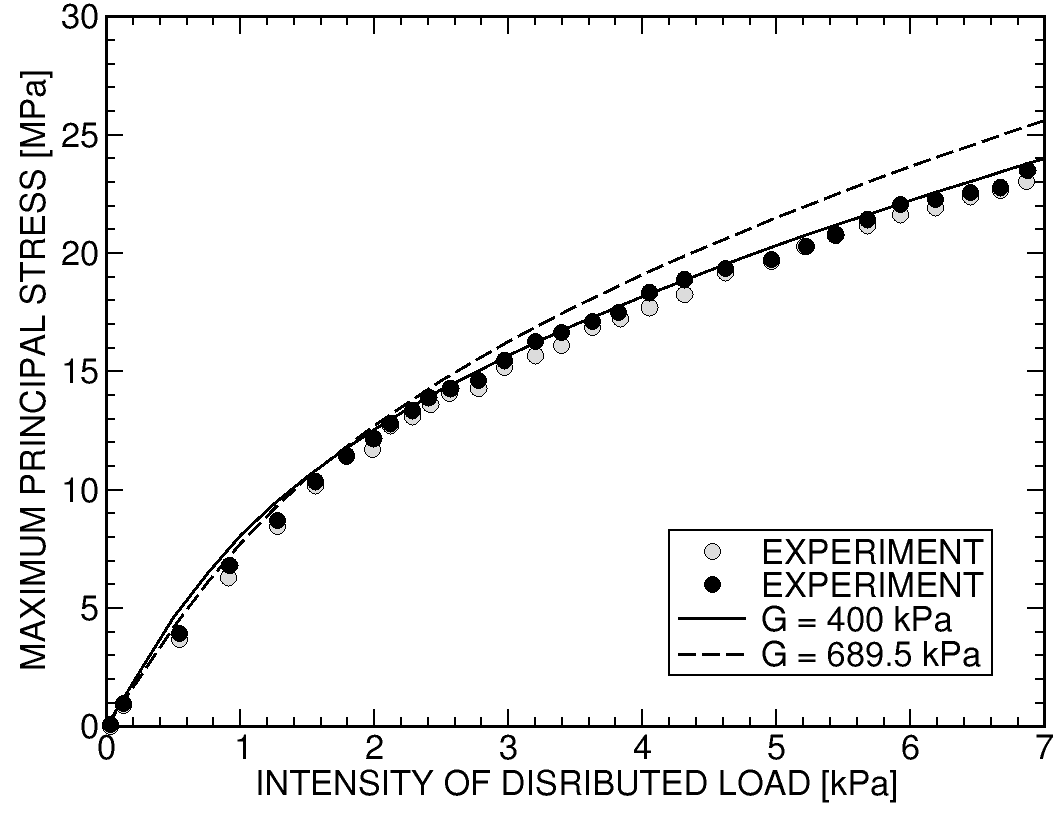} & \includegraphics[width=75mm]{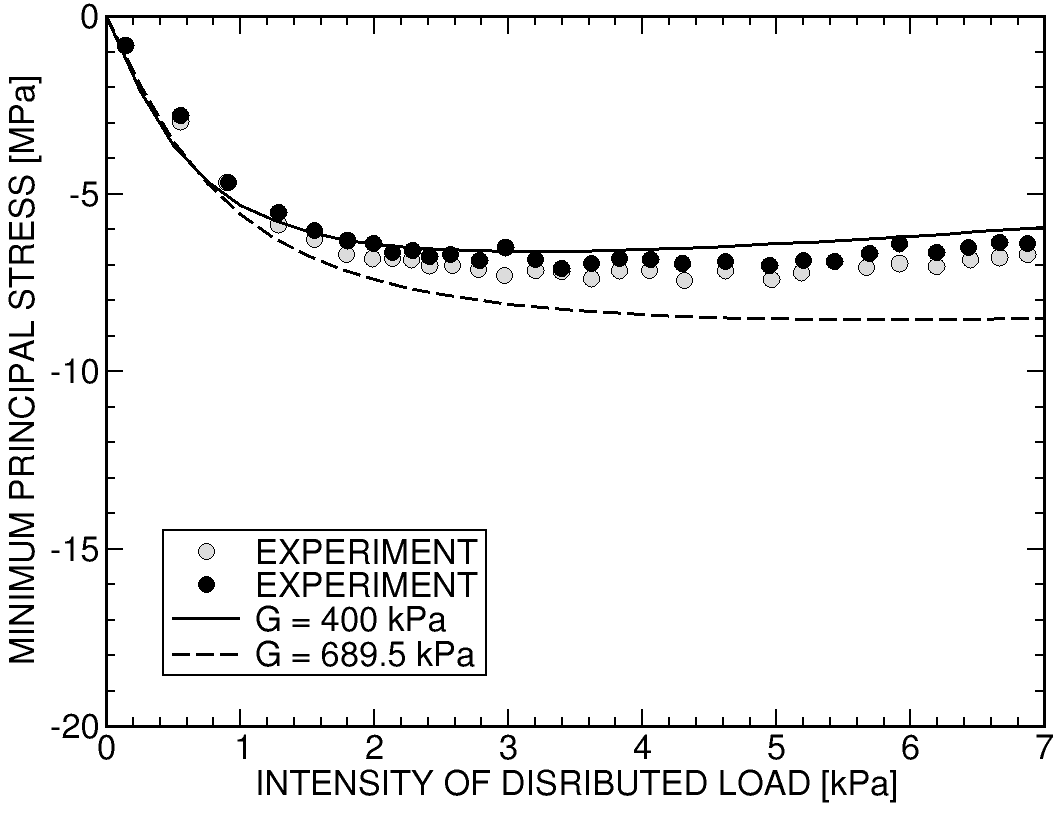}
\end{tabular}}
\caption{Comparison of experimental data for maximum principal stress at the center of the bottom surface of the bottom layer / for minimum principal stress at the center of the top surface of the top layer with the response of nonlinear finite element model of laminated glass plate with the PVB shear modulus 400~kPa or 689.5~kPa}
\label{fig:Exp_s1}
\end{figure}

\begin{figure}[p]
\centerline{%
\begin{tabular}{cc}
 \includegraphics[width=75mm]{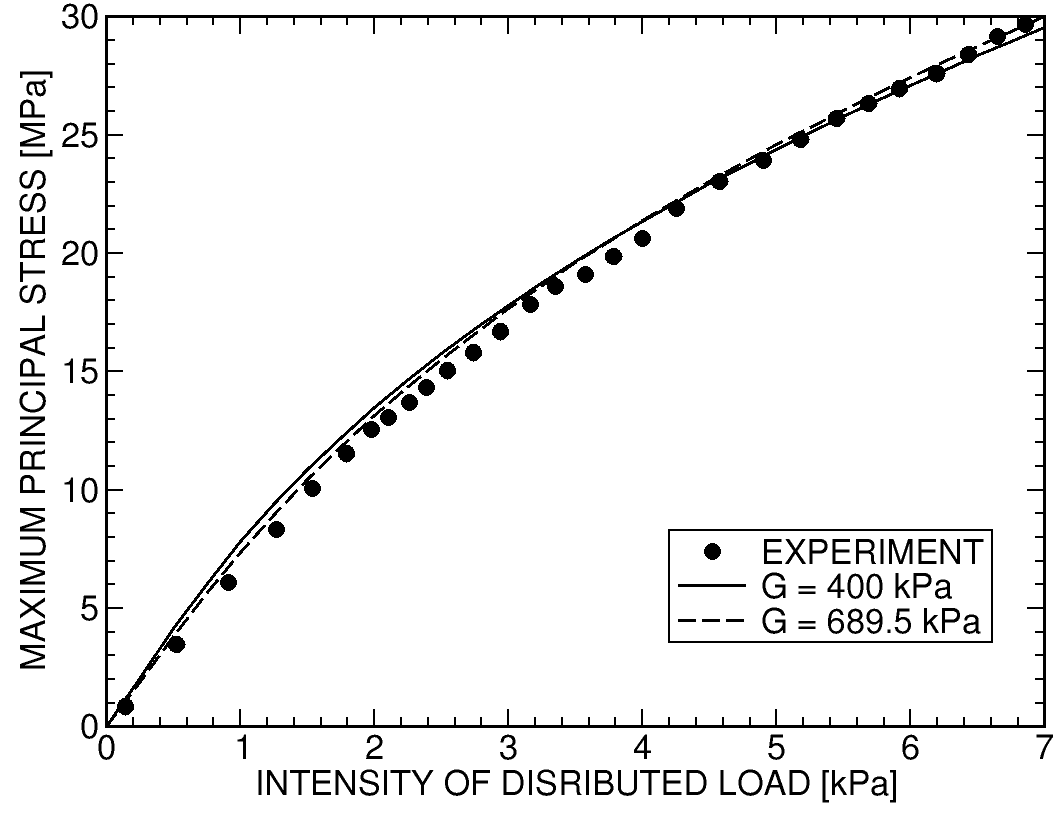} & \includegraphics[width=75mm]{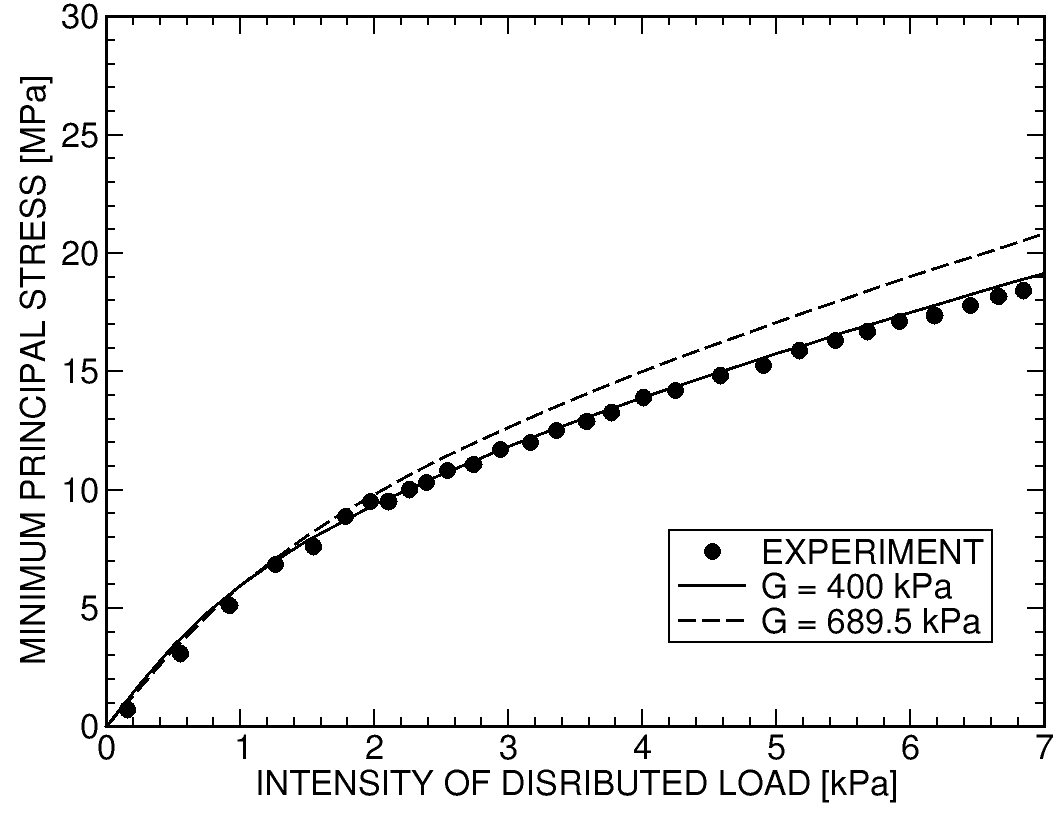}
\end{tabular}}
\caption{Comparison of experimental data for maximum and minimum principal stress at the point $[\frac{L_x}{4},\frac{L_y}{2}]$ of the bottom surface of the bottom layer with the response of nonlinear finite element model of laminated glass plate with the PVB shear modulus 400~kPa or 689.5~kPa}
\label{fig:Exp_s1L2}
\end{figure}

\begin{figure}[p]
\centerline{%
\begin{tabular}{cc}
 \includegraphics[width=75mm]{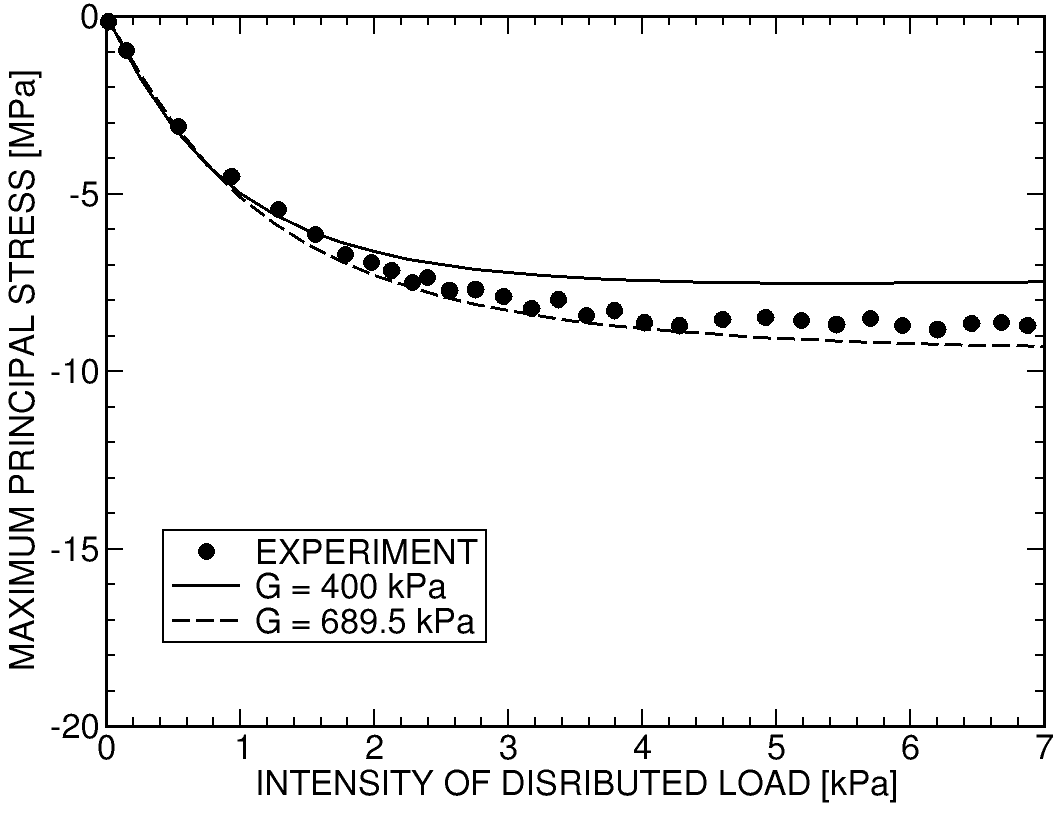} & \includegraphics[width=75mm]{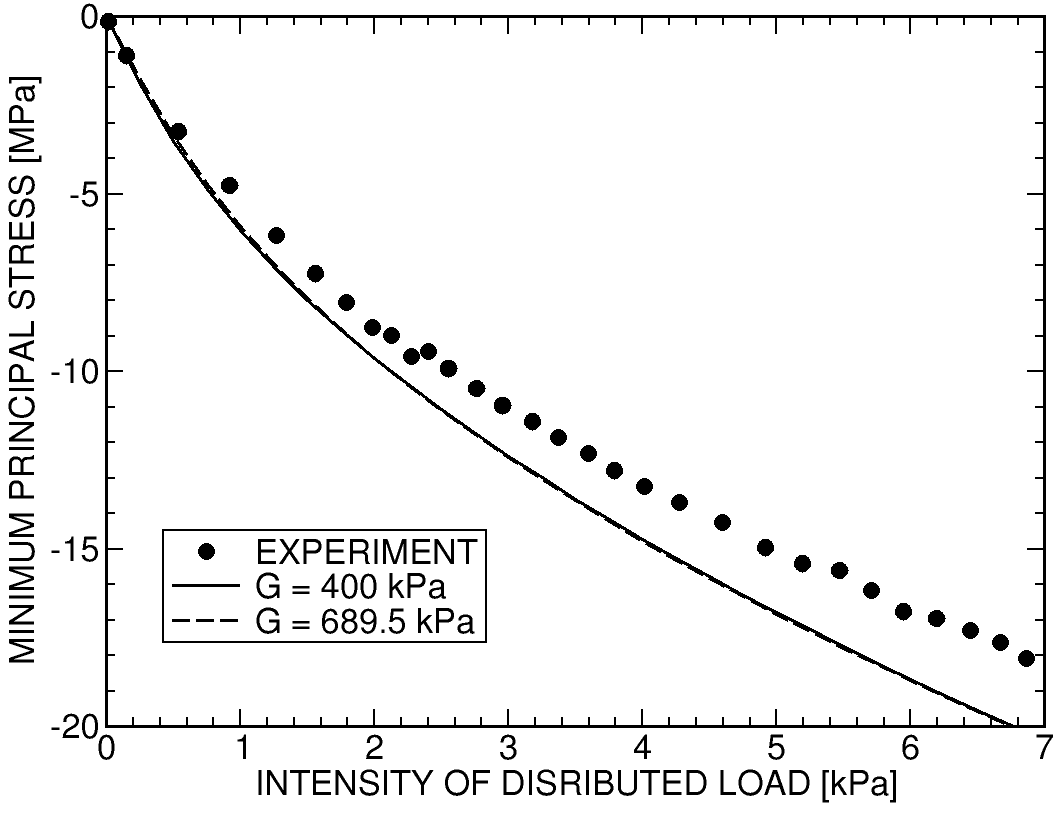}
\end{tabular}}
\caption{Comparison of experimental data for maximum and minimum principal stress at the point $[\frac{L_x}{4},\frac{L_y}{2}]$ of the top surface of the top layer with the response of nonlinear finite element model of laminated glass plate with the PVB shear modulus 400~kPa or 689.5~kPa}
\label{fig:Exp_s2L2}
\end{figure}

Due to the symmetry of the problem, we used a quarter of the plate for computations and discretized it with 50 $\times$ 50 elements per a layer. The dependence of the central deflection and principal stresses, at the center $[\frac{L_x}{2},\frac{L_y}{2}]$\footnote{The origin of the coordinate system is in the corner of the plate. Note that the two extreme principal stresses should be equal to each other at the center of the unit, because of symmetry. Therefore, two sets of experimental data appear in Figure~\ref{fig:Exp_s1}.} and at the point with coordinates $[\frac{L_x}{4},\frac{L_y}{2}]$, on the increasing intensity of distributed load $f_z$ appears in Figures~\ref{fig:Exp_w}--\ref{fig:Exp_s2L2}. We reproduced the experimental data from~\cite{Vallabhan:1993:ALG} and determined the principal stresses only from the membrane components in order to compare the principal stresses from the model directly to experimental values. The response of the finite element model was computed for two values of interlayer shear modulus, 400~kPa and 689.5~kPa. In addition, we included the prediction of the monolithic (thickness of two glass panes and interlayer together) and the layered glass plate (thickness of two independent glass panes), corresponding to the lower and upper bound for deflections, but not for the principal stresses at the center of the unit. Due to the effects of geometric nonlinearity, the location of the extreme of principal stresses is not at the center of the plate, but moves with an increasing load. Such phenomenon is best visible in Figure~\ref{fig:Exp_s1} for the minimum principal stresses, where the measured values become practically constant at the load range of $3$--$5$~kPa and then even slightly increase.

In Figure~\ref{fig:Exp_w}, we present the comparison of experimental data for the central deflection with the response of the nonlinear model of laminated glass plates. The proposed nonlinear model represents the overall behavior of the unit well. We reached the best agreement for the values of central deflections corresponding to the intensity of the distributed load between 3--5~kPa, which suggests that the constant value of the shear modulus $G\layer{2}$ is optimally adjusted to this interval. For lower values of $f_z$, the experimental values of central deflections are smaller and closer to the monolithic limit, for $f_z > 5$~kPa, the behavior of the unit is closer to the layered limit.

We attribute this phenomenon to the time-dependent behavior of the shear modulus of the interlayer under the constant temperature. The load duration, increasing with the value of the load intensity, corresponds to decreasing values of the effective shear modulus. Therefore, the effective value of the shear modulus for $f_z < 3$~kPa is higher than the optimized one, which results in lower values of the central deflections. The effective shear modulus of PVB is lower for $f_z > 5$~kPa, resulting in higher values of the central deflections. 
Thus, more accurate results can be obtained by viscoelastic analysis, but that requires the specification of the shear relaxation modulus of the interlayer, which is not available in the present case. %This suggests that the simplified elastic model appears to be sufficiently accurate.

Figure~\ref{fig:Exp_s1} shows the comparisons of the results for the maximum principal stresses at the center of the bottom surface of the bottom layer and for the minimum principal stresses at the center of the top surface of the top layer. We found excellent agreement between experimental and computed values for the maximum principal stresses and good agreement for the minimum principal stresses. Similar conclusions also follow from the comparison of extreme stresses at the top and bottom layers in a quarter of the plate, Figures~\ref{fig:Exp_s1L2} and~\ref{fig:Exp_s2L2}, in which we obtained very good agreement for tensile stresses and good agreement for compression stresses between the model predictions and experimental values. 

\begin{figure}[ht]
\centerline{%
\begin{tabular}{ccc}
\includegraphics[width=55mm]{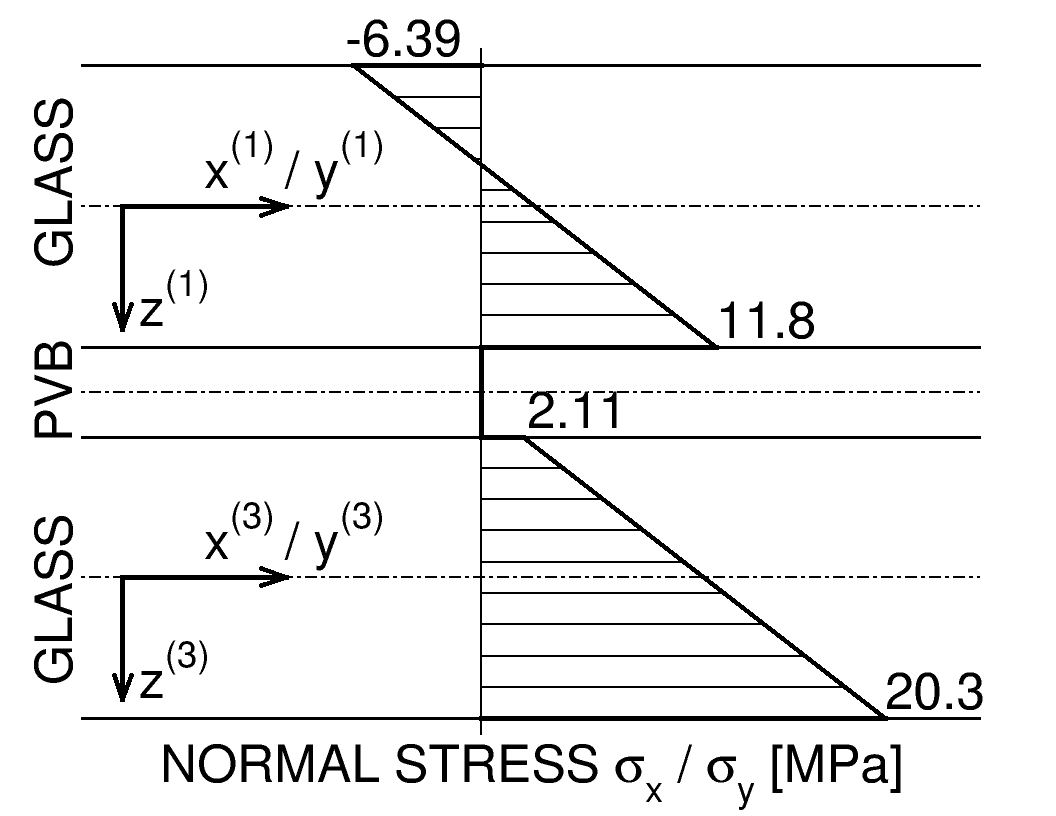} & \hspace{-7mm} \includegraphics[width=55mm]{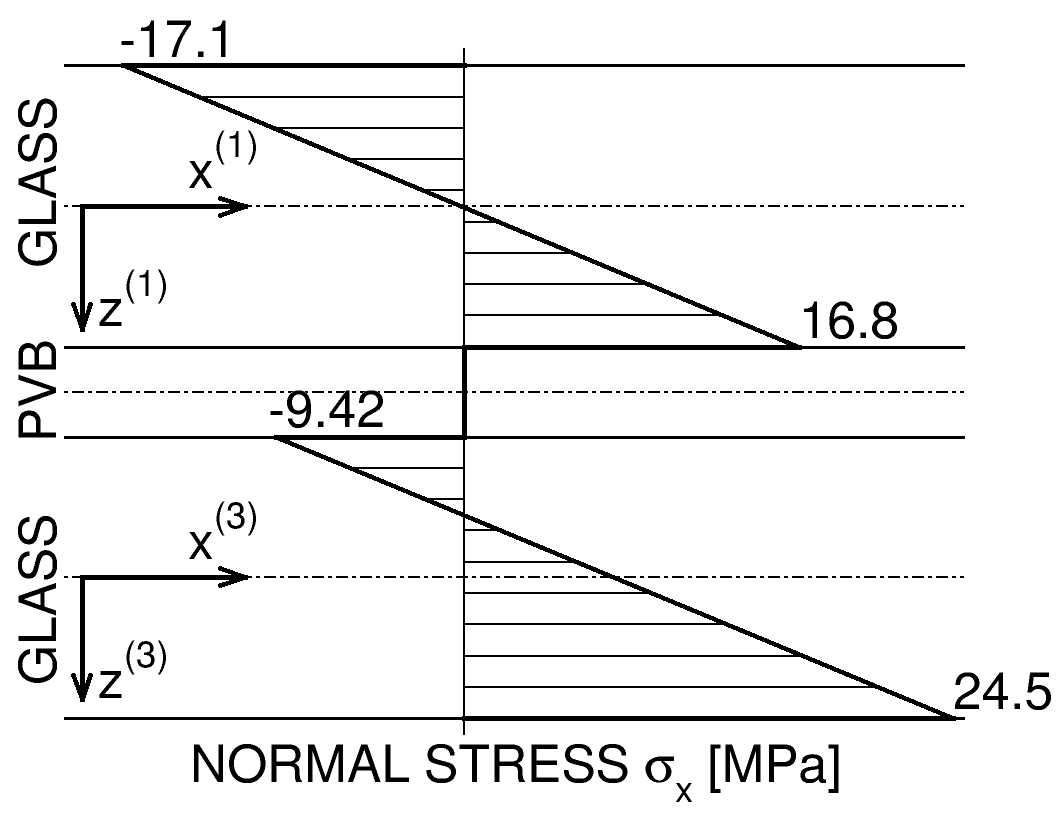} & \hspace{-5mm} \includegraphics[width=55mm]{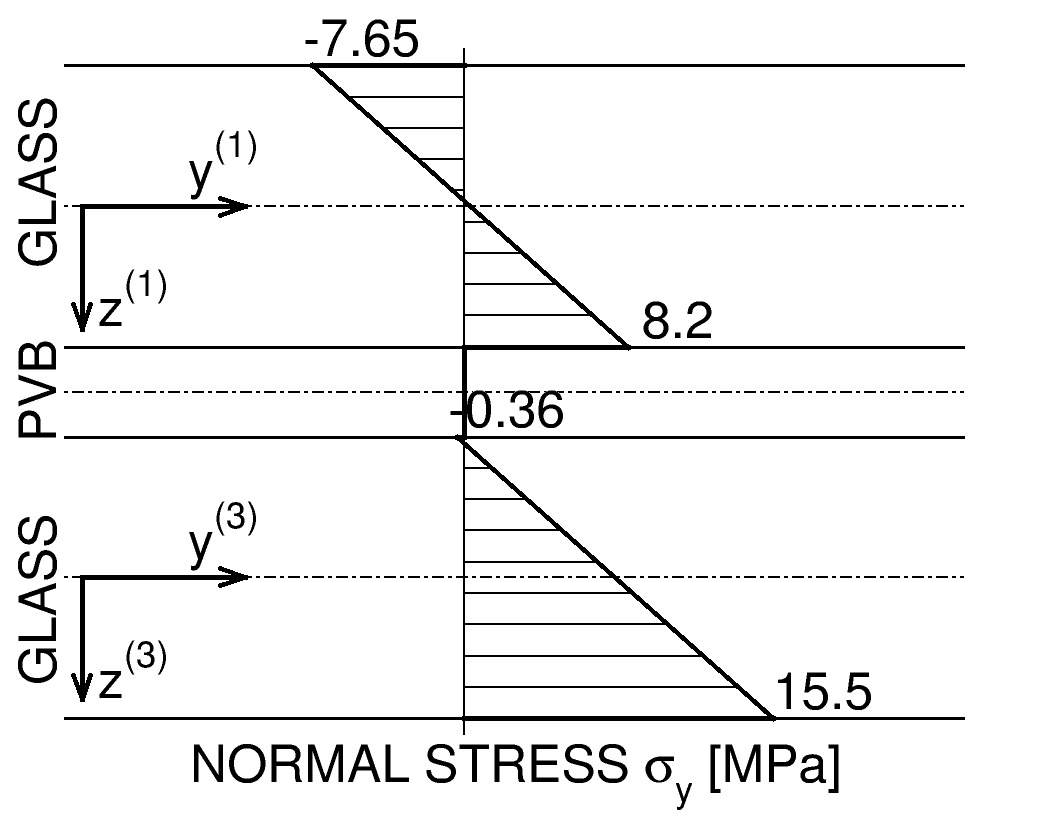} \\
at the center $[\frac{L_x}{2},\frac{L_y}{2}]$ & at the point $[\frac{L_x}{4},\frac{L_y}{2}]$ & at the point $[\frac{L_x}{4},\frac{L_y}{2}]$
\end{tabular}}
\caption{Distribution of normal stresses through the thickness determined by using the nonlinear finite element model for the simply supported squared plate uniformly loaded by~5~kPa}
\label{fig:FE_s}
\end{figure}

In Figure~\ref{fig:FE_s}, we plotted the distributions of normal stresses through the thickness of the previous simply supported laminated glass plate at the center $[\frac{L_x}{2},\frac{L_y}{2}]$ and at the point $[\frac{L_x}{4},\frac{L_y}{2}]$. The geometrically nonlinear behavior was assumed and the load intensity was set to 5~kPa. The normal stresses in the interlayer are negligible when compared with the normal stresses in the glass layers. The values of normal stresses at the mid-surfaces are generally nonzero, 
due to the presence of membrane stresses in the layers, and the load is equally distributed into the two glass layers, as visible from the identical slopes of the normal stresses. 

\begin{figure}[p]
\centerline{
\begin{tabular}{ccc}
\multicolumn{3}{c}{Maximum principal stresses on the bottom surface of the bottom layer} \\
\rotatebox{90}{\hspace{17mm}$f_z=$~1~kPa} &
 \includegraphics[height=48mm]{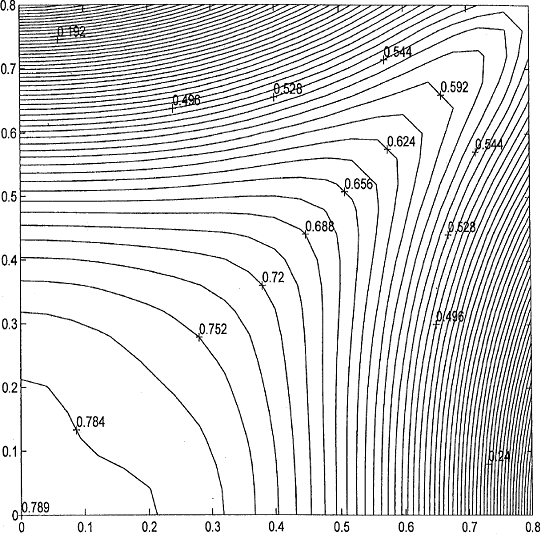} &
 \includegraphics[height=50mm]{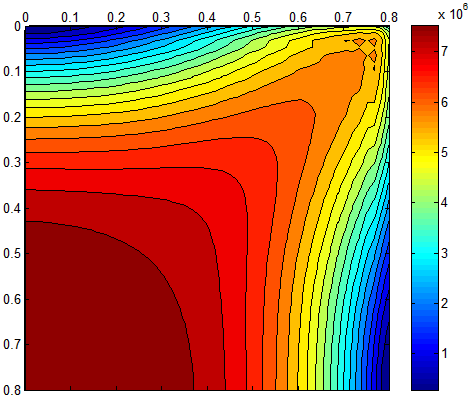} \\
 & 
(a) maximum 7.89~MPa & 
(b) maximum 7.89~MPa\\
 \rotatebox{90}{\hspace{16mm}$f_z=$~10~kPa} &
 \includegraphics[height=48mm]{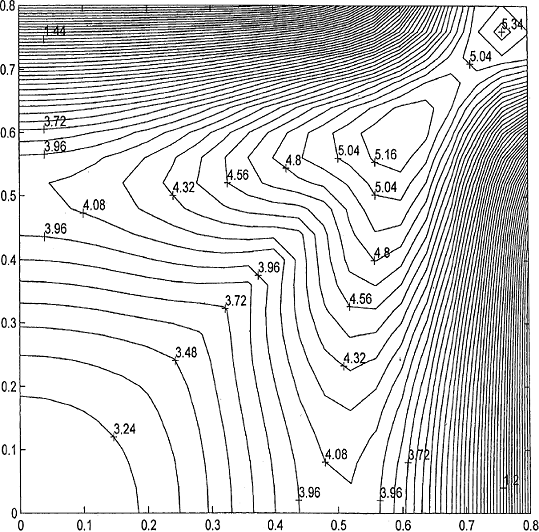} &
 \includegraphics[height=50mm]{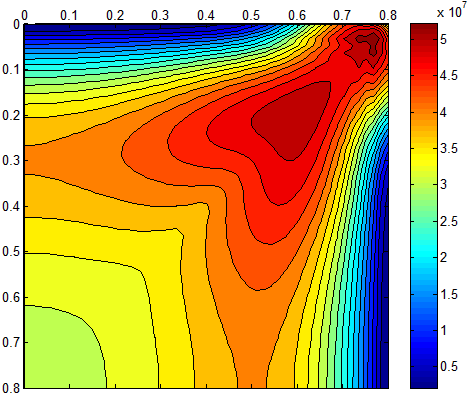}\\
 & 
(a) maximum 53.4~MPa & 
(b) maximum 54.7~MPa\\
\vspace{-1mm}\\
\multicolumn{3}{c}{Maximum principal stresses on the top surface of the top layer} \\
\rotatebox{90}{\hspace{17mm}$f_z=$~1~kPa} &
 \includegraphics[height=48mm]{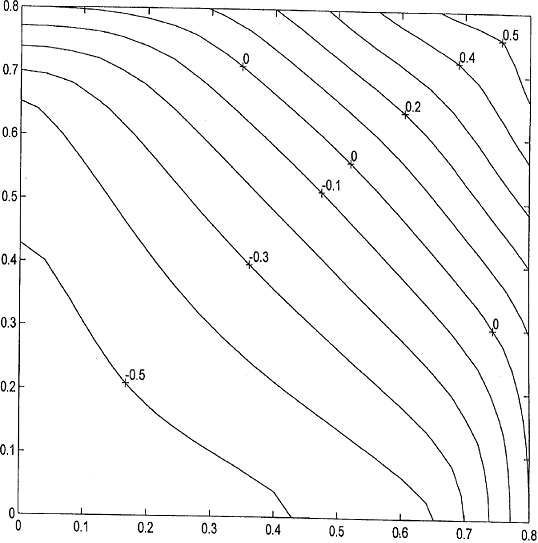} & 
 \includegraphics[height=50mm]{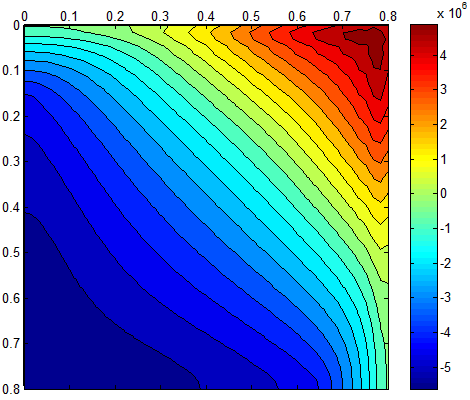} \\
 & 
(a) maximum 5~MPa & 
(b) maximum 5.43~MPa\\
\rotatebox{90}{\hspace{16mm}$f_z=$~10~kPa} &
\includegraphics[height=48mm]{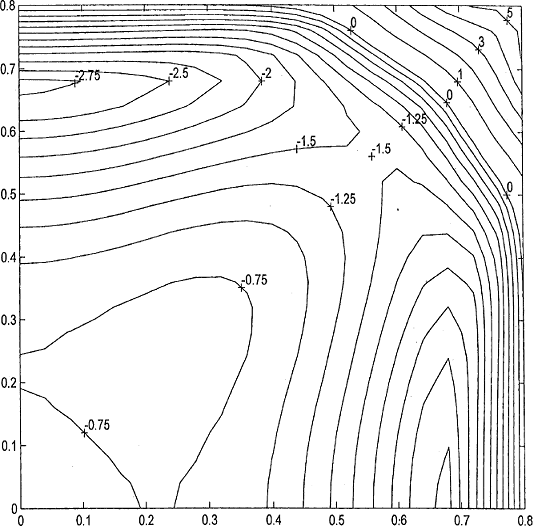} &
\includegraphics[height=50mm]{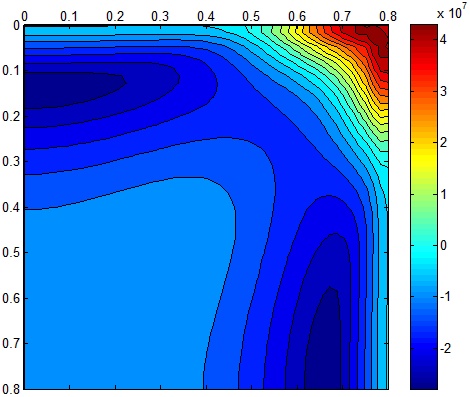}\\
 & 
(a) maximum 50~MPa & 
(b) maximum 46.7~MPa
\end{tabular}}
\caption{Contours of maximum principal stresses for distributed load of 
intensity $f_z$ obtained by (a) \cite{Asik:2003:LGP}, (b) nonlinear finite
element model of the laminated glass plate}
\label{fig:VV_stress}
\end{figure}

After validating the finite element model, we also present its verification against the finite difference solution by~\cite{Asik:2003:LGP}. The input data for the problem are the same as in~\cite{Vallabhan:1993:ALG}, except for slightly adjusted geometric parameters: the span of the plate is  1.6~m $\times$ 1.6~m and thicknesses of glass units equal to 5~mm. As shown in Figure~\ref{fig:VV_stress}, the distributions of extreme principal stresses on the top and bottom surfaces of the structure match very well. The largest difference in maximum values is around 8.5\%, which can be caused by only single-digit precision of the finite difference results for the top surface (the error due to this inaccuracy can be up to 10\%). The differences in maximum values for the bottom surface are under 2.5\%. Such an agreement is reasonable, given that the results were obtained by different numerical method and that the contours were constructed from different interpolations. 

\begin{figure}[ht]
\centerline{%
\begin{tabular}{cc}
 \includegraphics[width=75mm]{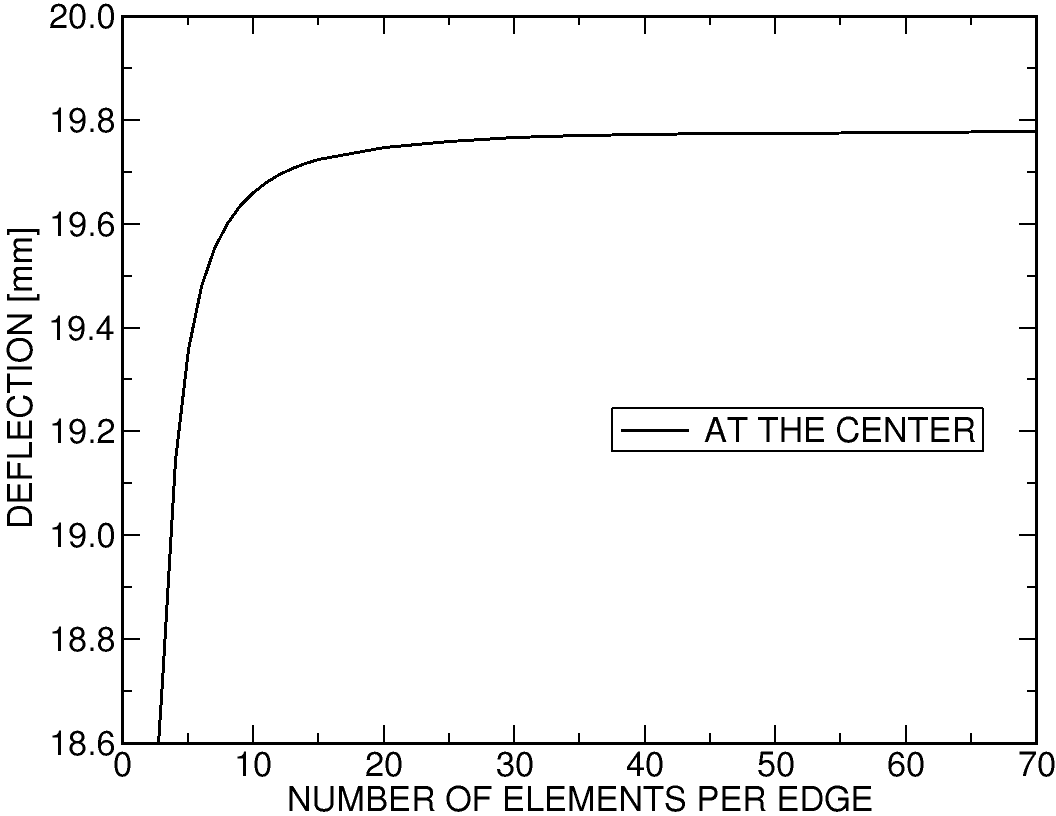} & \includegraphics[width=75mm]{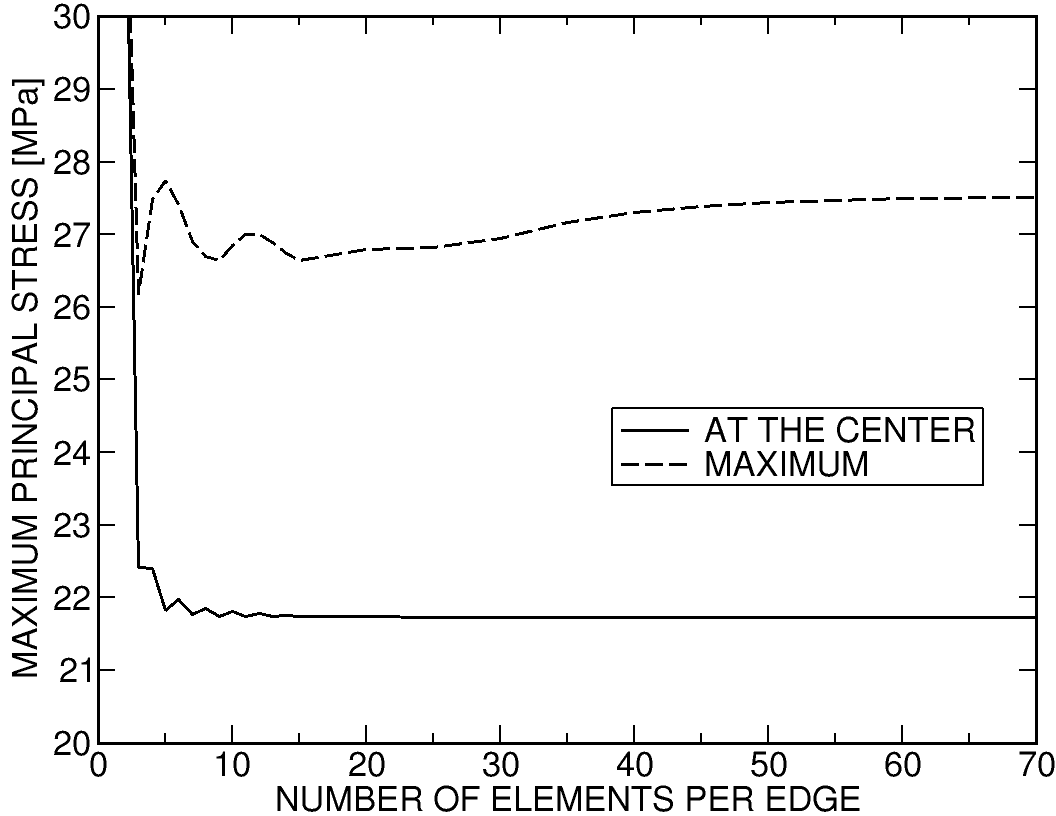}
\end{tabular}}
\caption{Convergence of nonlinear plate element upon uniform mesh refinement for $f_z=5$~kPa}
\label{fig:convergence_results}
\end{figure}

Finally, we also checked the convergence of response of the proposed model upon uniform mesh refinement. The dependence of the central deflections and the principal stress at the center and at the whole bottom layer appears in Figure~\ref{fig:convergence_results}. Both the central deflections and central extreme stresses exhibit a rapid convergence, while the accuracy of the largest principle stress found in the plate shows significant mesh-dependency. We attribute this phenomenon to the fact that, as a result of geometric nonlinearity, the extreme stresses become localized to a small area in the vicinity of the corner, Figure~\ref{fig:VV_stress}, where an adapted finite element mesh would be more suitable. Nevertheless, the discretization by $50\times 50$ elements appears to be sufficiently accurate for all the three quantities of interest.
%Finite element model~\cite{Duser:1999:AGBL}: description and verification. 

\subsection{Verification against semi-analytical model}\label{sec:verAM} 
%%%%%%%%%%%%%%%%%%%%%%%%%%%%%%%%%%%%%%%%%%%%%%%%%%%%%%%%%%%%%%%%%%%%%%%%%%%%%%%%%%%%%%%%%%%%%%%%%%%%%%%%%%%%%%%%%%%%%%%%%%%%%%%%%%
%
We analyzed again a simply supported square plate loaded with a uniformly distributed pressure 750~Pa. The dimensions of the plate were 3~m $\times$ 3~m, \Tref{Tab2:properties} summarize the thicknesses and material parameters of layers. The value of the shear modulus of the polymer interlayer was varied between $10^{-5}$~MPa and $10^{5}$~MPa.\footnote{Note that the value $10^{5}$~MPa is not realistic, but it is included in this study to assess the behavior of the analytical and finite element models in the true monolithic limit, in which the interlayer properties approach the properties of glass.} 

\begin{table}[ht]
\caption{Properties of glass layers and interlayer}
\centering{
\begin{tabular}{lcc}
\hline
\textbf{Layer/material} & \textbf{Glass} ($i=1,3$) & \textbf{Interlayer} ($i=2$)\\
\hline
Thickness $t\layer{i}$ [mm] & 10 & 1.52 \\
Young's modulus of elasticity $\E\layer{i}$ [MPa] & 70,000 & --\\
Shear modulus of elasticity $\G\layer{i}$ [MPa] & -- & $10^{-5}$ -- $10^{5}$ \\
Poisson's ratio $\nu\layer{i}$ [-] & 0.22 & 0.49\\
\hline
\end{tabular}
}
\label{Tab2:properties}
\end{table}

The details of the semi-analytical model for a simply-supported rectangular three-layered plate can be found in~\cite{Foraboschi:2012:AMLGP}. In short, the glass plies were assumed to behave as thin Kirchhoff plates under small deflections. The interlayer was supposed to be thin, compliant and to exhibit a state of pure shear. The resulting system was solved by the Fourier series expansion. 

 \begin{figure}[ht]
\centerline{%
\begin{tabular}{cc}
 \includegraphics[width=75mm]{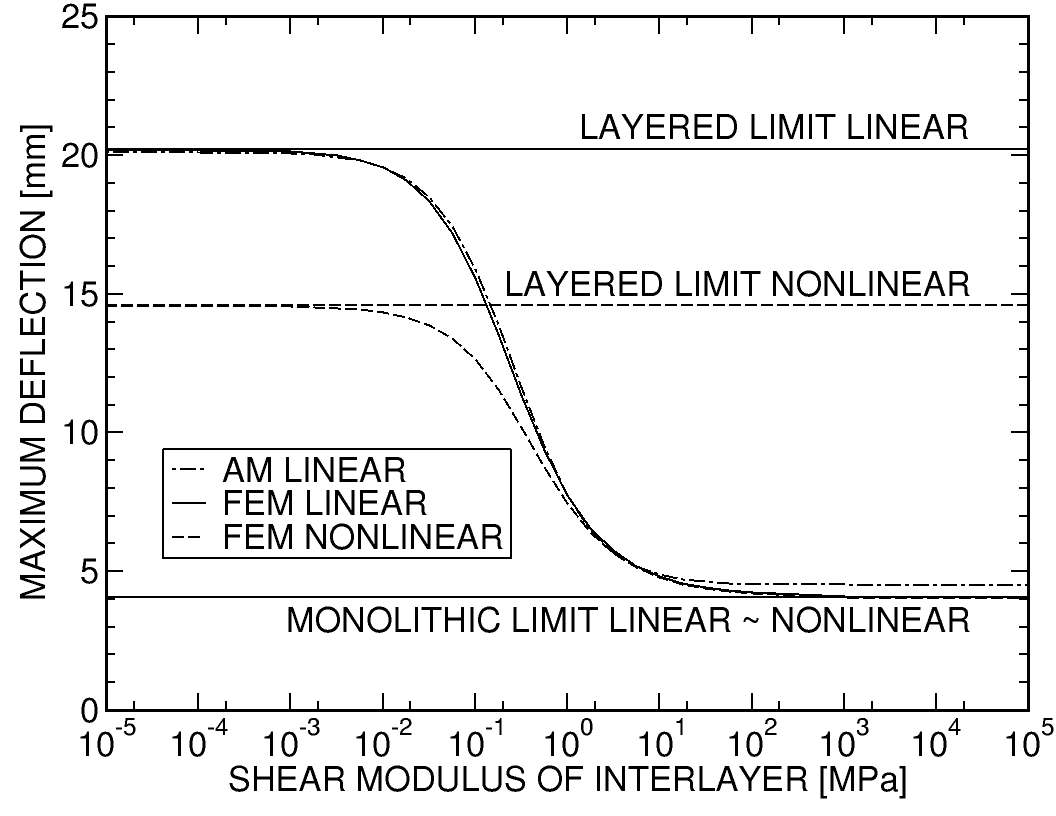} & \includegraphics[width=75mm]{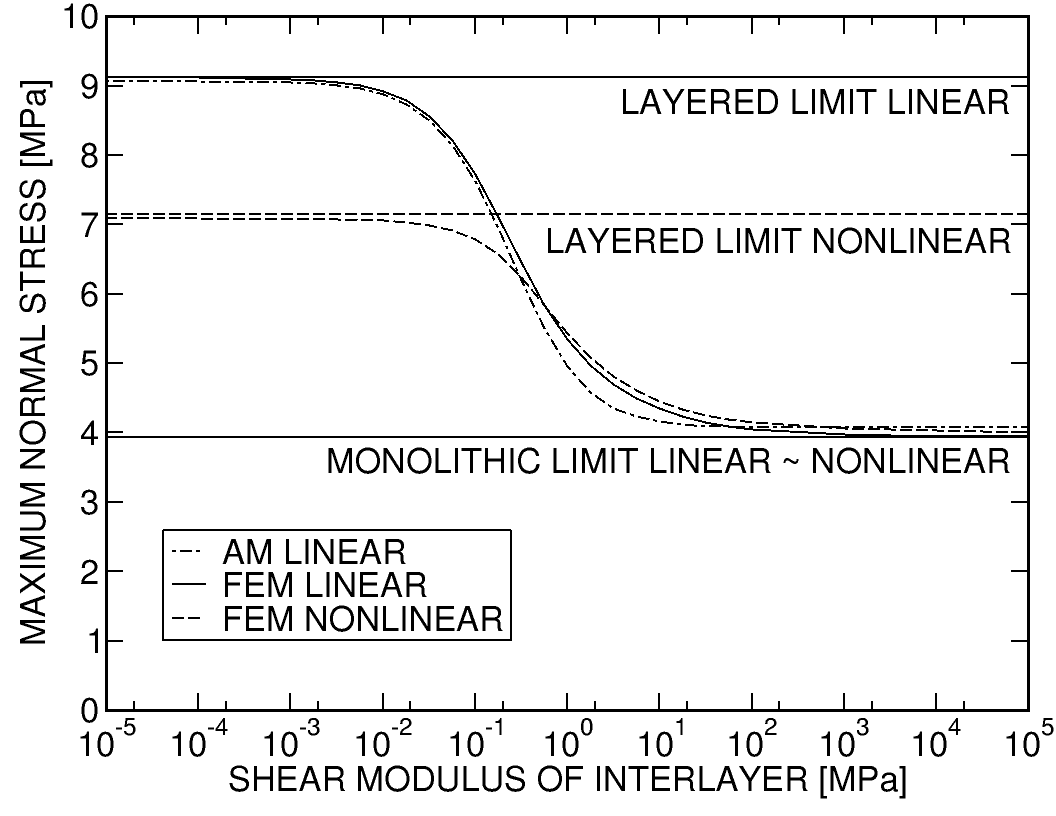}
\end{tabular}}
\caption{Comparison of maximum deflection and stress at the center of the simply supported plate obtained by semi-analytical solution (AM) and geometrical linear and nonlinear finite element model (FEM)}
\label{fig:AM_FEM}
\end{figure} 

The comparison of maximum deflection and normal stress for the semi-analytical and the finite element model appears in Figure~\ref{fig:AM_FEM}. We achieved an excellent agreement for deflections. The differences in values for the semi-analytical solution and geometrical linear finite element results stay under 2.5\% up to the shear modulus of 10~MPa. For larger values of the shear modulus, the deviations increase up to 10\%, because the semi-analytical model in~\cite{Foraboschi:2012:AMLGP} is not suitable for such relatively stiff interlayers,
for which the model does not approach the true monolithic response. The results for geometrically linear and nonlinear finite element model are almost identical for the shear modulus of interlayer greater than 1~MPa, but the differences increase up to 40\% for small values of the shear modulus (close to the layered limit). The maximum deflection at the center of plate for the geometrically nonlinear prediction ranges from $\frac{1}{600}$ to $\frac{1}{200}$ of the span, which implies that the effects of geometric nonlinearity can be significant in practically relevant cases.

Analogous conclusions follow for the normal stress distributions, Figure~\ref{fig:AM_FEM}. The agreement of solutions is good, the maximum error for geometrically linear model is about 8.5\%, which corresponds with conclusions in~\cite{Foraboschi:2012:AMLGP}. For the geometrically linear and nonlinear solution, the monolithic limit is almost the same, but the layered limit again shows important differences.

\subsection{Comparison with effective thickness approach}\label{sec:verET} 
\begin{figure}[p]
\vspace{-5mm}
\centerline{
\begin{tabular}{ccc}
%\multicolumn{2}{c}{Rectangular plate  with four edges simply supported} \\
\includegraphics[width=75mm]{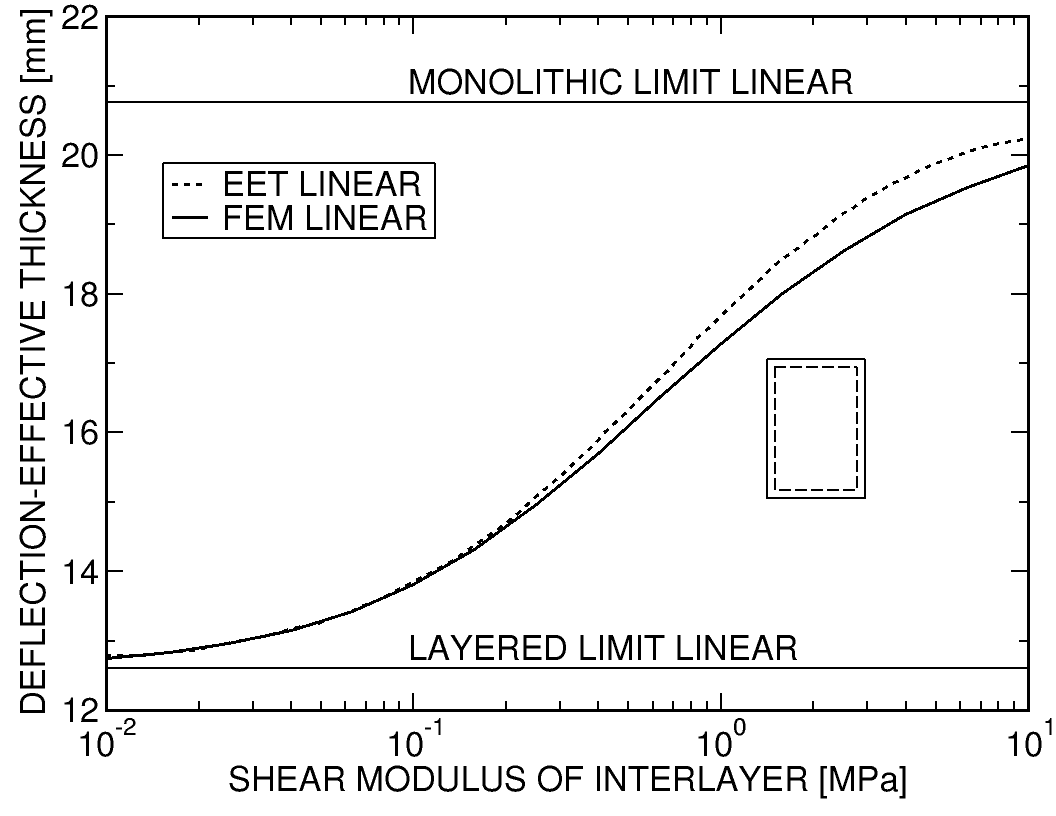} & \hspace{-5mm} (a) \hspace{-5mm} &  \includegraphics[width=75mm]{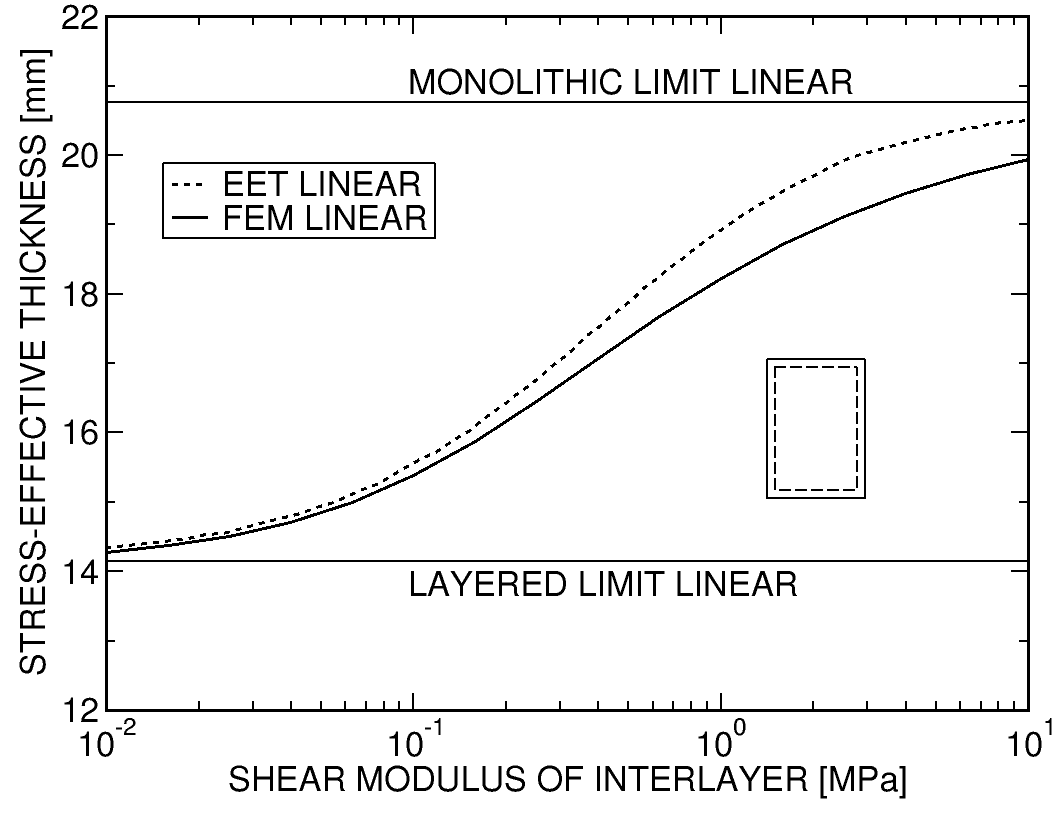} \\
%\multicolumn{2}{c}{Rectangular plate  with three edges simply supported} \\
\includegraphics[width=75mm]{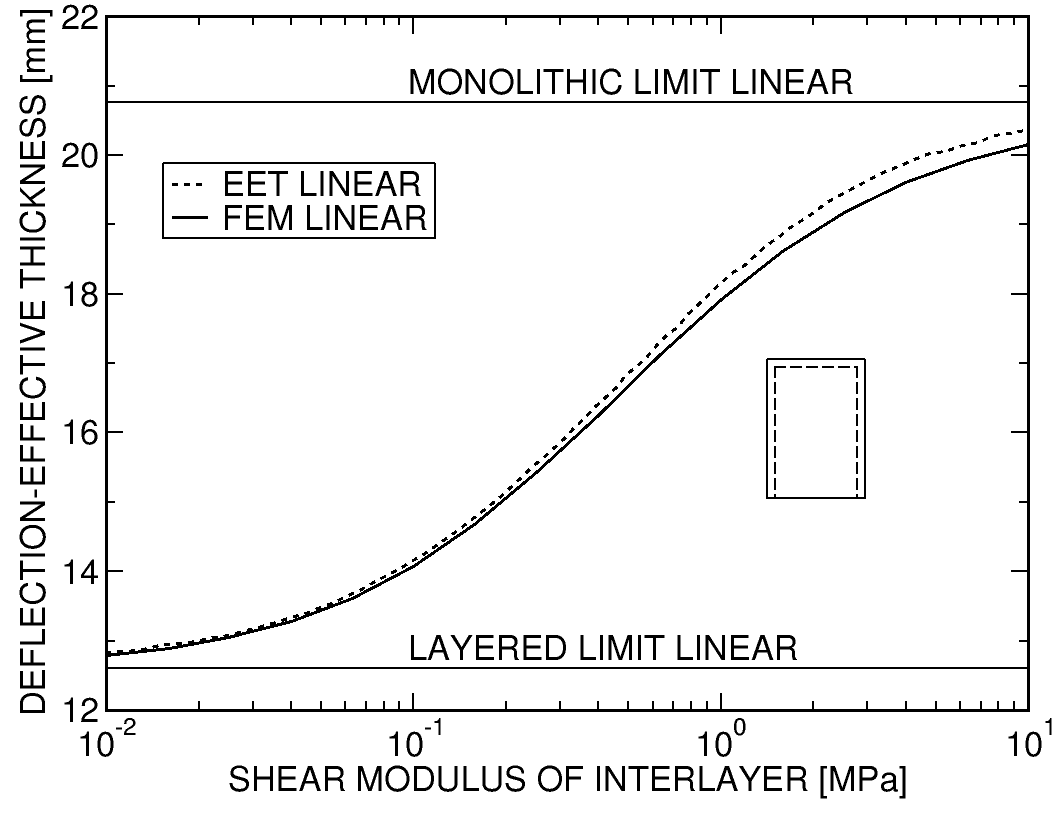} & \hspace{-5mm} (b) \hspace{-5mm} & \includegraphics[width=75mm]{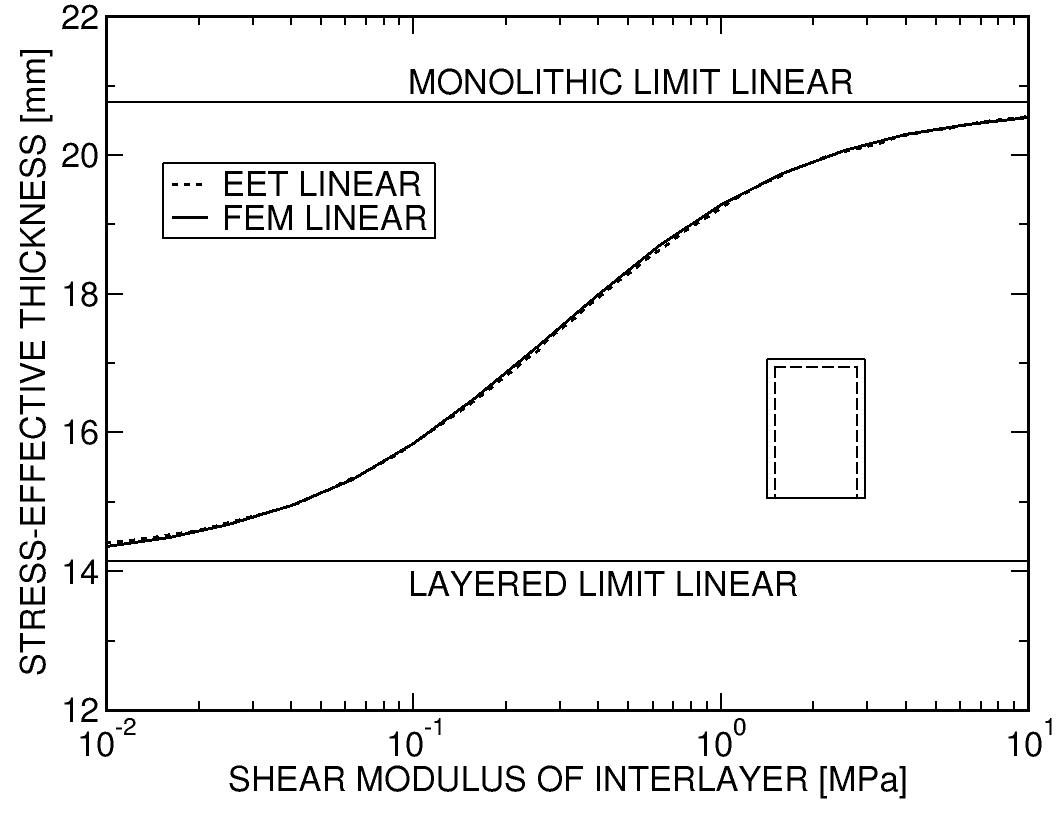} \\
%\multicolumn{2}{c}{Rectangular plate  with two opposite edges simply supported} \\
\includegraphics[width=75mm]{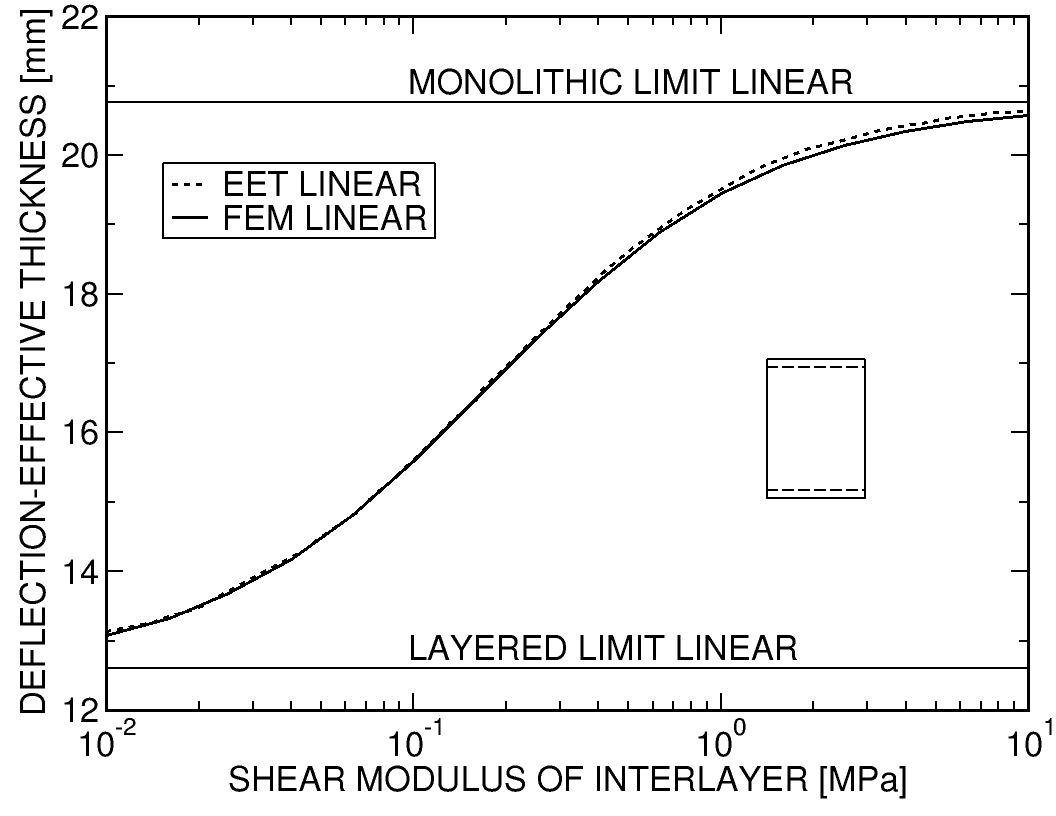} & \hspace{-5mm} (c) \hspace{-5mm} & \includegraphics[width=75mm]{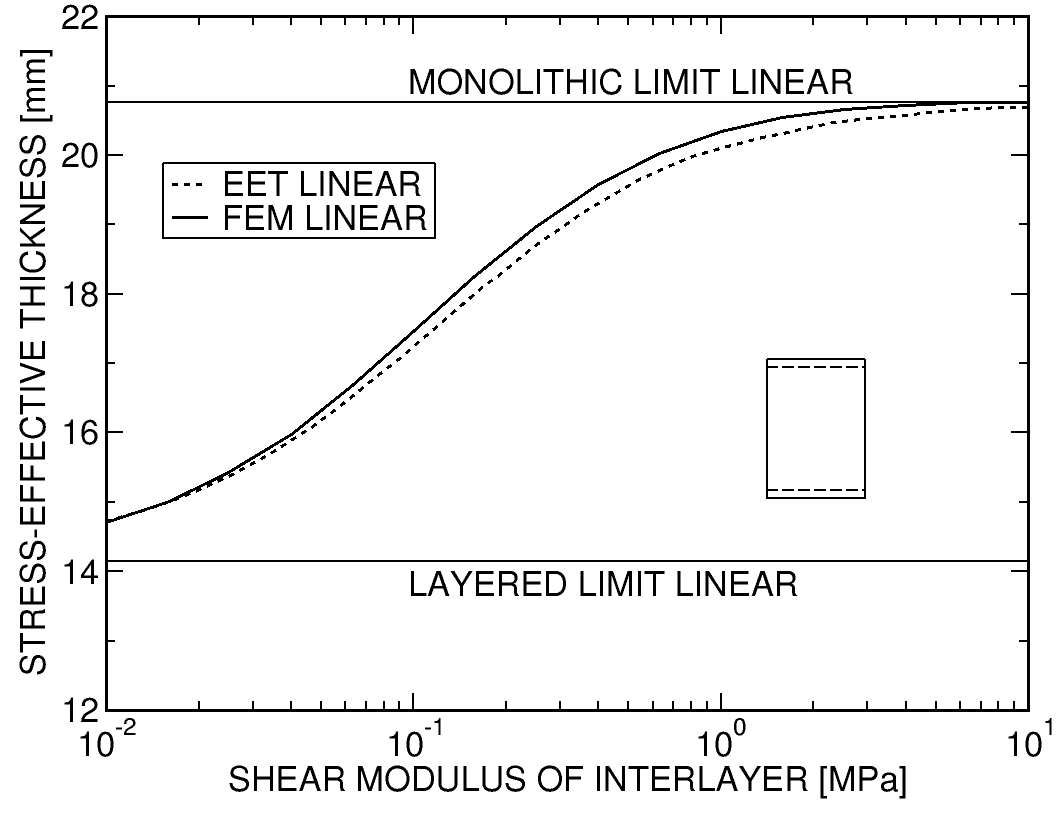} \\
%\multicolumn{2}{c}{Rectangular plate with two opposite edges simply supported and one built in} \\
\includegraphics[width=75mm]{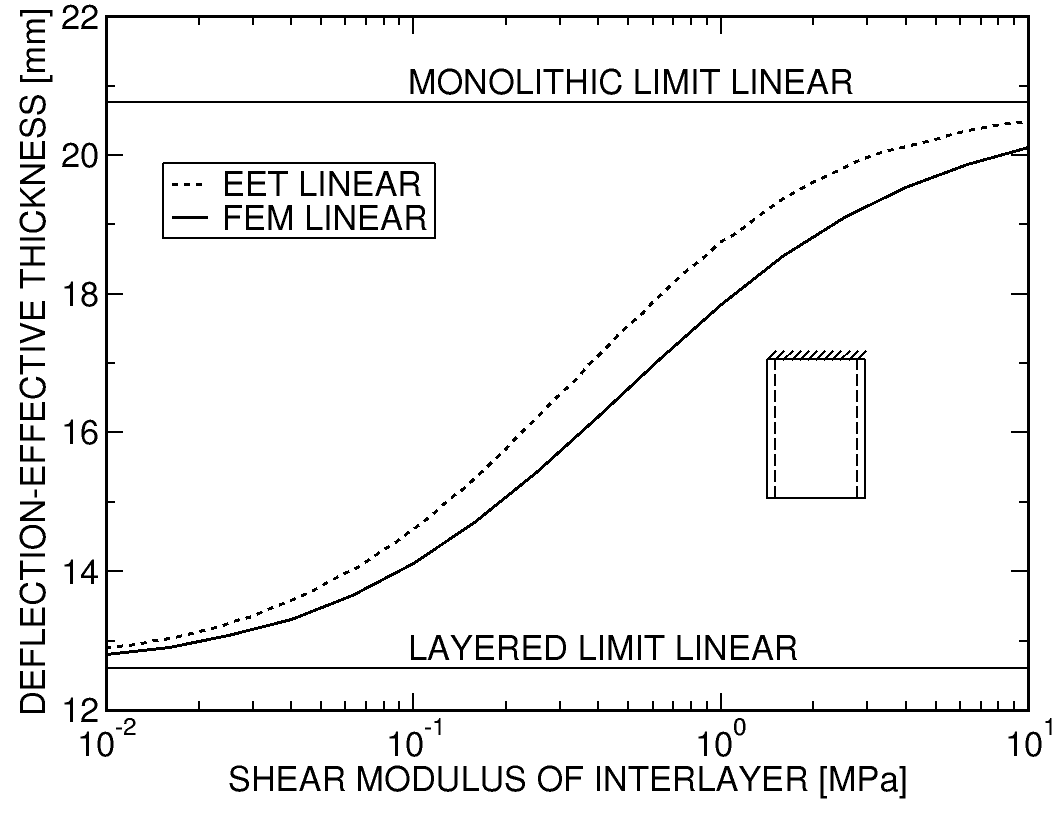} & \hspace{-5mm} (d) \hspace{-5mm} & \includegraphics[width=75mm]{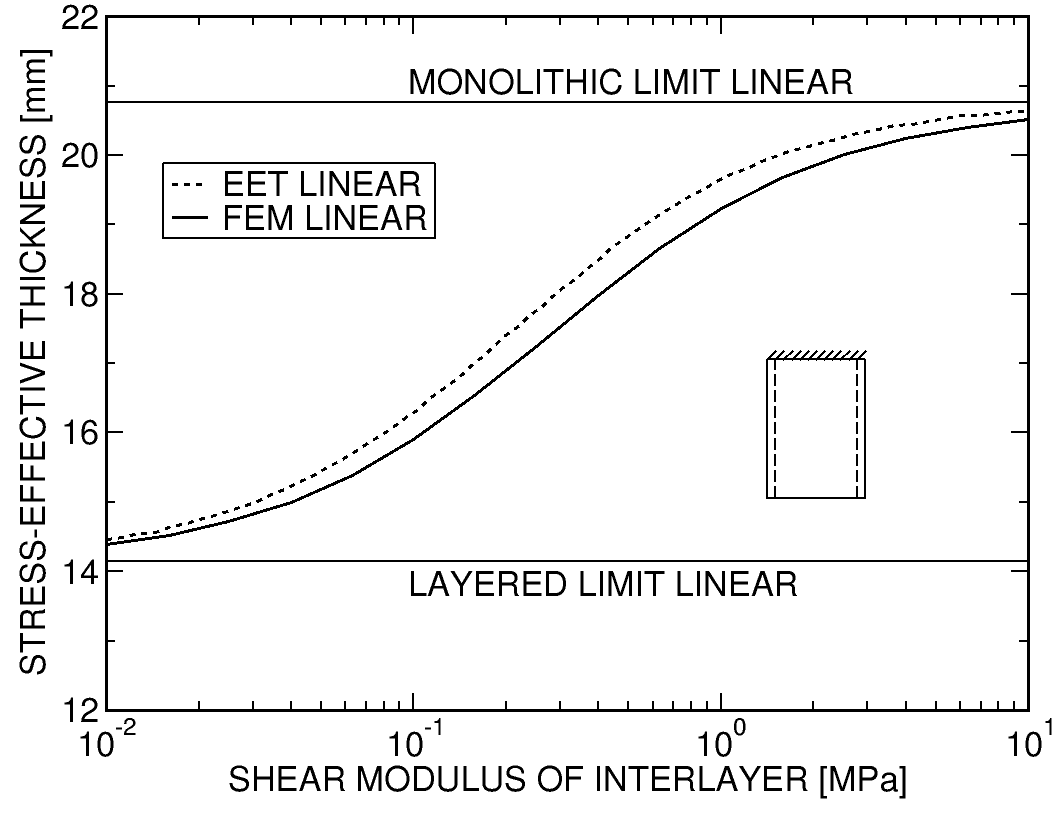} 
\end{tabular}}
\vspace{-2mm}
\caption{Comparison of the effective thickness obtained with the EET approach and computed from the response of the proposed model (FEM) for rectangular plates (a) with four edges simply supported, (b) with three edges simply supported, (c) with two opposite edges simply supported, (d) with two opposite edges simply supported and one fixed}
\label{fig:VV_EET}
\end{figure}

%%%%%%%%%%%%%%%%%%%%%%%%%%%%%%%%%%%%%%%%%%%%%%%%%%%%%%%%%%%%%%%%%%%%%%%%%%%%%%%%%%%%%%%%%%%%%%%%%%%%%%%%%%%%%%%%%%%%%%%%%%%%%%%%%%
%
One of the popular simplified approaches to the structural design of laminated glass units are effective thickness methods, e.g.~\cite{Benninson:2008:HPLG} or~\cite{Galuppi:2012:PEFD}. We compared the solution of our finite element model for laminated glass plates with the effective thickness approach using the enhanced effective thickness (EET) method of~\cite{Galuppi:2012:PEFD}, which gives more accurate results for deflection and stress calculations for different boundary conditions than the approach of~\cite{Benninson:2008:HPLG}. We  analyzed two rectangular plates loaded with a uniformly distributed pressure 750~Pa with the size 3~m $\times$ 2~m. \Tref{Tab3:properties} summarize the thicknesses and material parameters of layers. The value of shear modulus of the polymeric interlayer was varied between 0.01~MPa and 10~MPa. Four different boundary conditions were considered: four, three or two edges simply supported, and two opposite edges simply supported and one fixed, cf.~Figure~\ref{fig:VV_EET}.

\begin{table}[ht]
\caption{Properties of glass layers and interlayer}
\centering{
\begin{tabular}{lcc}
\hline
\textbf{Layer/material} & \textbf{Glass} ($i=1,3$) & \textbf{Interlayer} ($i=2$)\\
\hline
thickness $t\layer{i}$ [mm] & 10 & 0.76 \\
Young's modulus of elasticity $\E\layer{i}$ [MPa] & 70,000 & --\\
shear modulus of elasticity $\G\layer{i}$ [MPa] & -- & 0.01 -- 10 \\
Poisson's ratio $\nu\layer{i}$ [-] & 0.22 & 0.49\\
\hline
\end{tabular}
}
\label{Tab3:properties}
\end{table}

We compared the effective thicknesses computed according to~\cite{Galuppi:2012:PEFD} with the values obtained from the finite element modeling, Figure~\ref{fig:VV_EET}. The effective thickness method provides the deflection- and stress-effective thickness, which is the constant thickness of the equivalent monolithic homogeneous plate that has the same maximal deflection or maximal stress as laminated glass unit under the same boundary and load conditions. The effective thickness approach does not take into account the geometric nonlinearity, but in this example, the effect of geometric nonlinearity is minor. Therefore, we considered only the geometrically linear response of the finite element model. %We computed the maximum value of deflection and stress for laminated glass unit and found the thicknesses of equivalent monolithic homogenous plate. 
%The effective thicknesses were found by matching the maximum value of deflections and normal stresses for the finite element model (FEM) of the laminated glass unit to the response of the equivalent monolithic homogeneous plate. 
We found the effective thicknesses for the finite element model (FEM) by matching the maximum value of deflections and normal stresses of the laminated glass unit to the response of the equivalent monolithic homogeneous plate.
Due to the symmetry of the problem, we used only a quarter or a half of the plate for computations and discretized it with 40 $\times$ 60 or 40 $\times$ 120 elements per a layer, respectively. 

It can be observed that the enhanced effective thickness approach gives a very good approximations for the response of rectangular laminated glass subjected to a uniformly distributed transverse loading, which in practice coincide with the finite element solution for the plate with two or three edges simply supported. On the other hand, the enhanced effective method predicts effective thicknesses slightly greater than these from the numerical simulations for the other two boundary conditions, so that it underestimates the maximal deflections and stresses. 

\section{Conclusions}\label{sec:conclusions}
%%%%%%%%%%%%%%%%%%%%%%%%%%%%%%%%%%%%%%%%%%%%%%%%%%%%%%%%%%%%%%%%%%%%%%%%%%%%%%%%%%%%%%%%%%%%%%%%%%%%%%%%%%%%%%%%%%%%%%%%%%%%%%%%%%
% Conclusions and outlook

The assemblies of multiple elastic elements bonded by a compliant interlayer are frequently encountered in modern constructions.
The main contribution of this work is the development of a finite element model describing the laminated glass units based on the Mau refined theory for plates, which avoids fully resolved three-dimensional simulations, and its detailed verification and validation. We showed in the presented work that such an approach circumvents the limited accuracy of similar models available in most finite element systems, which employ a single set of kinematic variables and average the mechanical response through the thickness of the plate. %This is a consequence of heterogeneity of glass and interlayer material.

Although this paper was focused exclusively on the laminated glass units with two glass plates and an interlayer, the numerical models for multi-layered sandwich structures developed in this work are general and may find applications beyond glass structures. As a particular example, consider structural insulating panels consisting of a layer of polymeric foam sandwiched between two layers of structural board or laminated hybrid-glass units. %, see~\cite{Overend:2013:MPLHU}. 

This study represents a first step to the development of a comprehensive, mechanics-based model for laminated glass systems, which is suitable for
implementation in common engineering finite element solvers. As shown by the presented results, the proposed numerical methods are well-suited for the
modeling of laminated glass beams and plates, mainly because of the reduction of computational cost and accurate representation of the structural member
behavior. The other advantage of the proposed numerical framework is its conceptual simplicity, which will enable us to easily include additional
refinements of the model, such as a more accurate description of the time/temperature-dependence of the interlayer or delamination. %Note that the currently available simulation tool LaPla is designed for rectangular units, but the finite element formulation can be easily extended to different shapes of plates.

\paragraph{Acknowledgments:}
%%%%%%%%%%%%%%%%%%%%%%%%%%%%%%%%%%%%%%%%%%%%%%%%%%%%%%%%%%%%%%%%%%%%%%%%%%%%%%%%%%%%%%%%%%%%%%%%%%%%%%%%%%%%%%%%%%%%%%%%%%%%%%%%%%
\paragraph{}
This publication was supported by the European social fund within the framework of realizing the project {''Support of inter-sectoral mobility and quality enhancement of research teams at Czech Technical University in Prague'', CZ.1.07/2.3.00/30.0034.}
{Period of the project's realization 1.12.2012 -- 30.6.2015.} In addition, Jan Zeman acknowledges the partial support of the European Regional
Development Fund under the IT4Innovations Centre of Excellence, project No.~CZ.1.05/1.1.00/02.0070.

%The authors thank Professor Ji\v{r}\'{\i} \v{S}ejnoha of CTU in Prague for his helpfully comments on the original manuscript. 
%This work was supported by the , projects No.. 
%In addition, JZ acknowledges the partial support of the European Regional Development Fund under the IT4Innovations Centre of Excellence, project No.~CZ.1.05/1.1.00/02.0070.

\appendix
\section{Sensitivity analysis}\label{app:sensitivity_analysis}
%%%%%%%%%%%%%%%%%%%%%%%%%%%%%%%%%%%%%%%%%%%%%%%%%%%%%%%%%%%%%%%%%%%%%%%%%%%%%%%%%%%%%%%%%%%%%%%%%%%%%%%%%%%%%%%%%%%%%%%%%%%%%%%%%%
%\printnomenclature
%
We approximated the unknown fields of displacements and rotations by bi-linear functions at the four-node quadrilateral element
\begin{align}
\mtrx{u}\layer{i}_0(x,y) 
\approx
\M{N}\layer{i}\elen(x,y)
\mtrx{r}\layer{i}\elen,
&&
\mtrx{\rot}\layer{i}(x,y) 
\approx
\M{N}\layer{i}\eleb(x,y)
\mtrx{r}\layer{i}\eleb,
&&
\dispz_0\layer{i}(x,y)
\approx
\M{N}\layer{i}\elew(x,y)
\mtrx{r}\layer{i}\eles.
\end{align}
The vector $\mtrx{r}\layer{i}\el$  represents generalized nodal
displacements of the $e$-th element at the $i$-th layer and it will be convenient for further
discussion to organize the degrees of freedom into four vectors 
\begin{eqnarray*}
\mtrx{r}\layer{i}\elen & = & \left[ 
u\layer{i}\ele{1}, v\layer{i}\ele{1}, 
u\layer{i}\ele{2}, v\layer{i}\ele{2}, 
u\layer{i}\ele{3}, v\layer{i}\ele{3}, 
u\layer{i}\ele{4}, v\layer{i}\ele{4}
\right]\trn, \\
\mtrx{r}\layer{i}\eleb & = & \left[ 
\rot\layer{i}\elx{1}, \rot\layer{i}\ely{1}, 
\rot\layer{i}\elx{2}, \rot\layer{i}\ely{2}, 
\rot\layer{i}\elx{3}, \rot\layer{i}\ely{3}, 
\rot\layer{i}\elx{4}, \rot\layer{i}\ely{4}
\right]\trn, \\
\mtrx{r}\layer{i}\eles & = & \left[ 
w\layer{i}\ele{1}, \rot\layer{i}\elx{1}, \rot\layer{i}\ely{1}, 
w\layer{i}\ele{2}, \rot\layer{i}\elx{2}, \rot\layer{i}\ely{2}, 
w\layer{i}\ele{3}, \rot\layer{i}\elx{3}, \rot\layer{i}\ely{3}, 
w\layer{i}\ele{4}, \rot\layer{i}\elx{4}, \rot\layer{i}\ely{4}
\right]\trn,\\
\mtrx{r}\layer{i}\eleNL & = & \left[ 
w\layer{i}\ele{1}, w\layer{i}\ele{2}, w\layer{i}\ele{3}, w\layer{i}\ele{4}
\right]\trn,
\end{eqnarray*}
where the first three quantities correspond to the degrees of freedom contributing to the mid-surface membrane
strains, pseudo-curvatures, and transverse shear strains, respectively, and the last matrix $\mtrx{r}\layer{i}\eleNL$ stores the degrees of freedom
relevant to the nonlinear membrane strains, Eq.~\eqref{eq:strain_NL}. 

The matrices of basis function are formed in the standard way
\begin{eqnarray*}
\M{N}\layer{i}\elen(x,y)
=
\M{N}\layer{i}\eleb(x,y)
& = & 
\begin{bmatrix}
N_1 & 0 & N_2 & 0 & N_3 & 0 & N_4 & 0 \\
 0  & N_1 & 0 & N_2 & 0 & N_3 & 0 & N_4
\end{bmatrix},
\\
\M{N}\layer{i}\elew(x,y)
& = & 
\left[
\begin{array}{cccccccccccc}
N_1 & 0 & 0 & N_2 & 0 & 0 & N_3 & 0 & 0 & N_4 & 0 & 0 \\
\end{array}
\right].
\end{eqnarray*}

The matrices needed to evaluate the generalized strains then follow directly
from Eq.~\eqref{eq:generalized_strains} in the form
\begin{eqnarray*}
\M{B}\layer{i}\elen(x,y)
&=& 
 \pard \M{N}\layer{i}\elen(x,y),\\
\M{B}\layer{i}\eleb(x,y)
&=& 
 \pard \mtrx{S} \M{N}\layer{i}\eleb(x,y),\\
\M{B}\layer{i}\elew(x,y)
&=&
\M{\nabla}
\M{N}\layer{i}\elew(x,y),
\end{eqnarray*}
and the matrices for evaluation the von K\'{a}rm\'{a}n term in Eq.~\eqref{eq:generalized_strains_a} read as
\begin{eqnarray*}
\M{B}\layer{i}\eleNLx(x,y)
&=&
\begin{bmatrix}
\ppd{N_1}{x} & \ppd{N_2}{x} & \ppd{N_3}{x} & \ppd{N_4}{x} 
\end{bmatrix},\\
\M{B}\layer{i}\eleNLy(x,y)
&=&
\begin{bmatrix}
\ppd{N_1}{y} & \ppd{N_2}{y} & \ppd{N_3}{y} & \ppd{N_4}{y}
\end{bmatrix},
\end{eqnarray*}
and the matrix for evaluation the transverse shear strains from Eq.~\eqref{eq:generalized_strains_b}
\begin{eqnarray*}
\M{N}\layer{i}\eles(x,y)
&=& 
\left[
\begin{array}{cccccccccccc}
0 & 0 & N_1 & 0 & 0 & N_2 & 0 & 0 & N_3 & 0 & 0 & N_4 \\
0 & -N_1 & 0 & 0 & -N_2 & 0 & 0 & -N_3 & 0 & 0 & -N_4 & 0
\end{array}
\right].
\end{eqnarray*}

We used $2\times 2$ Gauss quadrature with four integration points $[x_g, y_g]^4_{g=1}$ to evaluate the normal and the
bending terms, and $1\times 1$ quadrature at the center for the shear terms as a remedy to shear locking. As a result, we obtained
\begin{align*}
\M{f}\layer{i}\ielen 
&=
\sum_{g=1}^4 \wf_g
\M{B}\layer{i}\elen\trn(x_g,y_g) 
\Dn\layer{i} 
\left(
  \M{B}\layer{i}\elen(x_g,y_g)\mtrx{r}\layer{i}\elen
  + 
  \mtrx{\strain}\mONLe\layer{i}\right)
\\
& = 
\sum_{g=1}^4 \wf_g
\M{B}\layer{i}\elen\trn(x_g,y_g) 
\M{n}\layer{i}\el(x_g, y_g),
\\
\M{f}\layer{i}\ieleb 
& = 
\sum_{g=1}^4 \wf_g
\M{B}\layer{i}\eleb\trn(x_g,y_g) 
\Db\layer{i} 
\M{B}\layer{i}\eleb(x_g,y_g)\mtrx{r}\layer{i}\eleb
=
\sum_{g=1}^4 \wf_g
\M{B}\layer{i}\eleb\trn(x_g,y_g) 
\M{m}\layer{i}\el(x_g, y_g),
\end{align*}
\begin{align*}
\M{f}\layer{i}\ieles 
& = 
\wf_0 
\left( 
  \M{B}\layer{i}\elew(x_0,y_0) + \M{N}\layer{i}\eles(x_0,y_0) 
\right)\trn
\Ds\layer{i} 
\left( 
  \M{B}\layer{i}\elew(x_0,y_0) + \M{N}\layer{i}\eles(x_0,y_0)
\right)
\mtrx{r}\layer{i}\eles
\\ 
& =
\wf_0
\left( 
  \M{B}\layer{i}\elew(x_0,y_0) + \M{N}\layer{i}\eles(x_0,y_0) 
\right)\trn
\M{q}\layer{i}\el(x_g, y_g),
\\
\M{f}\layer{i}\ieleNL 
& = 
\sum_{g=1}^4 \wf_g
\left(  \nabla%_{\mtrx{r}\layer{i}\eleNL} 
\mtrx{\strain}\mONLe\layer{i}(x_g,y_g)
\right)\trn \Dn\layer{i}
\left( \M{B}\layer{i}\elen(x_g,y_g)\mtrx{r}\layer{i}\elen +
\mtrx{\strain}\mONLe\layer{i}(x_g,y_g) \right)
\\
& = 
\sum_{g=1}^4 \wf_g
\left( \nabla%_{\mtrx{r}\layer{i}\eleNL} 
\mtrx{\strain}\mONLe\layer{i}(x_g,y_g)
\right)\trn
\M{n}\layer{i}\el(x_g, y_g),
\end{align*}
in which the additional terms read 
\begin{eqnarray*}
\mtrx{\strain}\mONLe\layer{i} (x_g,y_g) &=& 
\begin{bmatrix}
\frac{1}{2}\left( \M{B}\layer{i}\eleNLx(x_g,y_g) \mtrx{r}\layer{i}\eleNL \right)^2\\
\frac{1}{2}\left( \M{B}\layer{i}\eleNLy(x_g,y_g) \mtrx{r}\layer{i}\eleNL \right)^2\\
\M{B}\layer{i}\eleNLx(x_g,y_g) \mtrx{r}\layer{i}\elenNL \M{B}\layer{i}\eleNLy(x_g,y_g) \mtrx{r}\layer{i}\eleNL
\end{bmatrix},\\
\nabla\mtrx{\strain}\mONLe\layer{i}(x_g,y_g) &=& 
\tiny{
\begin{bmatrix}
\left( \M{B}\layer{i}\eleNLx(x_g,y_g) \mtrx{r}\layer{i}\eleNL \right) \M{B}\layer{i}\eleNLx(x_g,y_g)\\
\left( \M{B}\layer{i}\eleNLy(x_g,y_g) \mtrx{r}\layer{i}\eleNL \right) \M{B}\layer{i}\eleNLy(x_g,y_g)\\
\left( \M{B}\layer{i}\eleNLx(x_g,y_g) \mtrx{r}\layer{i}\eleNL \right) \M{B}\layer{i}\eleNLy(x_g,y_g) +
\left( \M{B}\layer{i}\eleNLy(x_g,y_g) \mtrx{r}\layer{i}\eleNL \right) \M{B}\layer{i}\eleNLx(x_g,y_g)
\end{bmatrix}},
\end{eqnarray*}
and $\wf_g$ and $\wf_0$ are weight functions for four- or one-point numerical integration.

Differentiating the terms $\M{f}\layer{i}\ielen$, $\M{f}\layer{i}\ieleb$, $\M{f}\layer{i}\ieles$, and $\M{f}\layer{i}\ieleNL$ with respect to the corresponding degrees of freedom
produces the linear and nonlinear parts of the stiffness matrices. Thus, the linear contributions to the element
stiffness matrix become
\begin{eqnarray*}
\M{\K}\layer{i}\elen &=& \sum_{g=1}^4 \wf_g\M{B}\layer{i}\elen\trn(x_g,y_g) \Dn\layer{i} \M{B}\layer{i}\elen(x_g,y_g),\\
\M{\K}\layer{i}\eleb &=& \sum_{g=1}^4 \wf_g\M{B}\layer{i}\eleb\trn(x_g,y_g) \Db\layer{i} \M{B}\layer{i}\eleb(x_g,y_g),\\
\M{\K}\layer{i}\eles &=& \wf_0\left( \M{B}\layer{i}\elew(x_0,y_0) + \M{N}\layer{i}\eles(x_0,y_0) \right)\trn
\Ds\layer{i} \left( \M{B}\layer{i}\elew(x_0,y_0) + \M{N}\layer{i}\eles(x_0,y_0) \right).
\end{eqnarray*}
The treatment of the ``von K\'{a}rm\'{a}n term'' is more involved, due to the nonlinear interaction of quantities
$\mtrx{r}\layer{i}\elen$ and $\mtrx{r}\layer{i}\eleNL$. We expressed the additional contributions to the stiffness matrix in the form
\begin{align*}
\M{\K}\layer{i}\elenNL &= \sum_{g=1}^4 \wf_g
\M{B}\layer{i}\elen\trn(x_g,y_g) \Dn\layer{i} \nabla\mtrx{\strain}\mONLe\layer{i}(x_g,y_g),\\
\M{\K}\layer{i}\eleNLn &= \M{\K}\layer{i}\elenNL \trn,
\\
\M{\K}\layer{i}\eleNL &= \sum_{g=1}^4 \wf_g
\left( \nabla\mtrx{\strain}\mONLe\layer{i}(x_g,y_g) \right) \trn
\Dn\layer{i}
\nabla\mtrx{\strain}\mONLe\layer{i}(x_g,y_g)\\
&+
\sum_{g=1}^4 \wf_g
\begin{bmatrix}
\M{\K}\layer{i}\eleNLc{1} &
\M{\K}\layer{i}\eleNLc{2} &
\M{\K}\layer{i}\eleNLc{3} &
\M{\K}\layer{i}\eleNLc{4}
\end{bmatrix},
\end{align*}
with
\begin{eqnarray*}
\M{\K}\layer{i}\eleNLg
 =
\begin{bmatrix}
{B}\layer{i}\eleNLxg(x_g,y_g) \M{B}\layer{i}\eleNLx\trn(x_g,y_g)\\
{B}\layer{i}\eleNLyg(x_g,y_g) \M{B}\layer{i}\eleNLy\trn(x_g,y_g)\\
{B}\layer{i}\eleNLxg(x_g,y_g) \M{B}\layer{i}\eleNLy\trn(x_g,y_g) +
{B}\layer{i}\eleNLyg(x_g,y_g) \M{B}\layer{i}\eleNLx\trn(x_g,y_g)
\end{bmatrix}
\M{n}\layer{i}\el(x_g, y_g).
\end{eqnarray*}
Note that, for example, the matrix $\M{\K}\layer{i}\elenNL$ is localized to rows corresponding to the degrees of freedom collected in $\mtrx{r}\layer{i}\elen$ and the columns corresponding to $\mtrx{r}\layer{i}\eleNL$, and ${B}\layer{i}\eleNLxg(x_g,y_g)$ abbreviates the $c$-th entry of matrix $\M{B}\layer{i}\eleNLx$. 

% We obtained the response of the whole structure by standard element assembly operations, e.g.~\citep{Bittnar:1996:NMM}.

\end{document}

%% file: format.tex
\newcommand{\de}[1]{\,{\mathrm d}#1}

\newcommand{\Tref}[1]{Table~\ref{#1}}

\newcommand{\mtrx}[1]{{\boldsymbol{#1}}} % A matrix
\newcommand{\trn}{{\sf ^T}}        % Transpose
\newcommand{\pard}{\mtrx{\partial}}

\newcommand{\dispx}{u} % x displacement
\newcommand{\dispy}{v} % y displacement
\newcommand{\dispz}{w} % z displacement

\newcommand{\rotx}{\varphi_x} % Rotation around the x axis
\newcommand{\roty}{\varphi_y} % Rotation arouth the y axis

\newcommand{\curv}{\kappa} % Curvature
\newcommand{\curvx}{\kappa_x}
\newcommand{\curvy}{\kappa_y}
\newcommand{\curvxy}{\kappa_{xy}}

\newcommand{\normal}{n}
\newcommand{\shear}{q}
\newcommand{\bendy}{m}

\newcommand{\rot}{\varphi}

\newcommand{\strain}{\varepsilon}
\newcommand{\sstrain}{\gamma}

\newcommand{\m}{_m}
\newcommand{\mO}{_{m0}}
\newcommand{\mONL}{_{m0,NL}}
\newcommand{\mONLe}{_{m0,NL,e}}
\newcommand{\Dm}{\mtrx{D}_m}
\newcommand{\Dn}{\mtrx{D}_n}
\newcommand{\Db}{\mtrx{D}_b}
\newcommand{\Ds}{\mtrx{D}_s}
\newcommand{\h}{h}
\newcommand{\correct}{k}
\newcommand{\G}{G}
\newcommand{\K}{K}
\newcommand{\E}{E}

\newcommand{\stress}{\sigma}
\newcommand{\sstress}{\tau}

\newcommand{\del}[2]{\mbox{$\displaystyle\frac{#1}{#2}$}}
\newcommand{\ppd}[2]{\del{\partial {#1}}{\partial{#2}}}

\newcommand{\layer}[1]{^{(#1)}}
\newcommand{\lay}[1]{^{(#1)}}
\newcommand{\el}{_e}
\newcommand{\ele}[1]{_{e,#1}}
\newcommand{\elx}[1]{_{x,e,#1}}
\newcommand{\ely}[1]{_{y,e,#1}}

\newcommand{\elen}{_{n,e}}
\newcommand{\eleb}{_{b,e}}
\newcommand{\eles}{_{s,e}}
\newcommand{\eleNL}{_{K,e}}
\newcommand{\eleNLg}{_{Kw,e,c}}
\newcommand{\eleNLc}[1]{_{Kw,e,#1}}
\newcommand{\eleNLx}{_{K,x,e}}
\newcommand{\eleNLy}{_{K,y,e}}
\newcommand{\eleNLxg}{_{K,x,e,c}}
\newcommand{\eleNLyg}{_{K,y,e,c}}
\newcommand{\elenNL}{_{nK,e}}
\newcommand{\eleNLn}{_{Kn,e}}
\newcommand{\ielen}{_{\mathrm{int},n,e}}
\newcommand{\ieleb}{_{\mathrm{int},b,e}}
\newcommand{\ieles}{_{\mathrm{int},s,e}}

\newcommand{\ieleNL}{_{\mathrm{int},K,e}}

\newcommand{\eln}[1]{_{e,#1}}
\newcommand{\nel}{{N\el}}

\newcommand{\Eint}{\Pi_{\mathrm{int}}}
\newcommand{\Eext}{\Pi_{\mathrm{ext}}}
\newcommand{\Etot}{\Pi}
\newcommand{\Eexte}[2]{\Etot\lay{#1}_{\mathrm{ext},#2}}
\newcommand{\Einte}[2]{\Etot\lay{#1}_{\mathrm{int},#2}}

\newcommand{\dmn}{\Omega}

\newcommand{\M}[1]{{\boldsymbol #1}}

\newcommand{\Mfint}{\M{f}_\mathrm{int}}
\newcommand{\Mfext}{\M{f}_{\mathrm{ext}}}
\newcommand{\Mfinte}[1]{\M{f}_{\mathrm{int},#1}}
\newcommand{\Mfexte}[2]{\M{f}_{\mathrm{ext},#2}\lay{#1}}

\newcommand{\ite}[1]{{^{#1}}}
\newcommand{\I}{\mtrx{I}}

\newcommand{\Md}{\M{r}}
\newcommand{\Mde}[2]{\Md^{(#1)}_{#2}}
\newcommand{\MKt}{\M{K}}%_\mathrm{t}}
\newcommand{\Eref}[1]{Eq.~\eqref{#1}}

\newcommand{\wf}{\alpha}
\newcommand{\step}{k}